%% file: main.tex
\documentclass[twocolumn,preprintnumbers,superscriptaddress,aps,nofootinbib]{revtex4-2}
\usepackage{amsmath,amssymb,mathtools}
\usepackage{graphicx}
\usepackage{booktabs,cellspace}
\usepackage{color}
\usepackage{lineno}
\usepackage{caption}
\usepackage{subcaption}
\usepackage{subcaption}
\usepackage{hyperref}
\usepackage{xspace}
\usepackage[normalem]{ulem}
\usepackage{soul}
\usepackage{mathrsfs}
\usepackage{placeins}
\usepackage{float}
\usepackage{tikz-feynman}
\tikzfeynmanset{compat=1.1.0}

\makeatletter 
\renewcommand\onecolumngrid{
\do@columngrid{one}{\@ne}%
\def\set@footnotewidth{\onecolumngrid}
\def\footnoterule{\kern-6pt\hrule width 1.5in\kern6pt}%
}
\makeatother
\usepackage{ragged2e}
\makeatletter
\let\original@makecaption\@makecaption
\def\@makecaption#1#2{%
  \original@makecaption{#1}{\justifying #2\par}%
}
\makeatother
\newcommand{\as}{\ensuremath{\alpha_s}\xspace}

\newcommand{\pb}{\ensuremath{\,\mathrm{pb}}\xspace}

\newcommand{\pth}{\ensuremath{p_T^H}\xspace}
\newcommand{\ptw}{\ensuremath{p_T^W}\xspace}

\newcommand{\qtcut}{\ensuremath{q_T^{\rm cut}}\xspace}

\usepackage{pdfcomment}

\newcommand{\IPPP}{Institute for Particle Physics Phenomenology,
  Department of Physics, University of Durham, Durham, DH1 3LE, UK}
\newcommand{\ETH}{Institute for Theoretical Physics, ETH, 8093 Z\"{u}rich, Switzerland}
\newcommand{\UZH}{Physik-Institut, Universit\"at Z\"urich,
  Winterthurerstrasse 190, 8057 Z\"urich, Switzerland}
\newcommand{\Torino}{Dipartimento di Fisica,
  Universit\`a degli Studi di Torino\\and INFN, Sezione di Torino, Via P. Giuria 1,
  10125, Torino, Italy}
\newcommand{\CERNaff}{CERN, Theoretical Physics Department, CH-1211
 Geneva 23, Switzerland}
\newcommand{\Bicocca}{Dipartimento di Fisica G. Occhialini, U2,
  Universit\`a degli Studi di Milano-Bicocca and INFN, Sezione di
  Milano-Bicocca, Piazza della Scienza, 3, 20126 Milano, Italy}

\begin{document}
%


\title{Boosted Higgs-strahlung off a $W$ boson \\at next-to-next-to-next-to-leading order in QCD}

\preprint{CERN-TH-2026-114, IPPP/26/44, LAPTH-034/26,  ZU-TH 21/26}

\author{Aude Gehrmann-De Ridder} \affiliation{\ETH}\affiliation{\UZH}
\author{Alexander Huss} \affiliation{\CERNaff}  %
\author{Matteo Marcoli} \affiliation{\IPPP}
\author{Pier Francesco Monni} \affiliation{\CERNaff}  %
\author{Emanuele Re} \affiliation{\Bicocca}
\author{Luca Rottoli} \affiliation{\Bicocca}
\author{Federico Silvetti} \affiliation{\IPPP}
\author{Paolo Torrielli} \affiliation{\Torino}

\begin{abstract}
  \noindent
  The production of a boosted Higgs boson in association with a
  charged weak ($W$) boson is a key process to scrutinize the
  electroweak symmetry breaking mechanism at hadron colliders. This
  reaction constitutes the dominant Higgs production channel at large
  transverse momentum, providing unique sensitivity to Higgs-boson
  interactions with other Standard Model particles as well as to physics
  beyond the Standard Model.
  In this Letter, we present the first fully differential calculation
  of this important scattering process at
  next-to-next-to-next-to-leading order (N$^3$LO) in perturbative Quantum
  Chromodynamics (QCD).
  We find that the N$^3$LO corrections, amounting to approximately $+2\%$ in the boosted regime, generally lie at the edge of or outside the standard scale variation band of the previous perturbative order. The residual dependence of the N$^3$LO prediction on perturbative scales is reduced to below the percent level, marking a milestone for the Higgs precision program.
\end{abstract}

\maketitle
\paragraph*{Introduction. ---}
The associated production of a Higgs boson with a charged weak gauge boson (Higgs-strahlung) plays a central role in the precision program of the Large Hadron Collider (LHC) \cite{ATLAS:2020fcp, ATLAS:2021tbi, ATLAS:2023jdk, CMS:2020zge, CMS:2021nnc, CMS:2024ddc}.
In particular, the boosted regime, characterized by large transverse momenta of the Higgs boson, provides enhanced sensitivity to the Higgs couplings. This, in turn, enables powerful indirect probes of physics beyond the Standard Model~\cite{ Baglio:2020oqu, Bizon:2016wgr, Bizon:2021rww, Gauld:2023gtb, Bonetti:2025hnb}. For such a kinematic selection, Higgs-strahlung becomes one of the dominant Higgs production modes~\cite{Becker:2020rjp}, and for this reason the study of this regime will play a key role in the Higgs program in the High-Luminosity phase of the LHC (HL-LHC), where the high-transverse-momentum tail will be probed precisely.

From a theoretical perspective, Higgs-strahlung also constitutes a benchmark process for precision QCD calculations. Inclusive predictions for the total cross section are known up to next-to-next-to-next-to-leading order (N$^3$LO) in perturbative QCD~\cite{Baglio:2022wzu}, while electroweak (EW) corrections are currently known up to NLO~\cite{Ciccolini:2003jy,Denner:2011id,Denner:2014cla,Granata:2017iod}. However, at the differential level, the available accuracy for the production cross section is currently limited to NNLO~\cite{Ferrera:2011bk,Ferrera:2013yga,Astill:2016hpa,Campbell:2016jau,Caola:2017xuq,Ferrera:2017zex,Alioli:2019qzz,Behring:2020uzq, Majer:2020kdg, Zanoli:2021iyp, Gauld:2019yng, Haisch:2022nwz, Gauld:2023gtb}. The increase in experimental precision foreseen at HL-LHC demands theoretical predictions at higher orders, especially in fiducial regions of phase space.

In this Letter, we present the first computation of Higgs-strahlung off a $W^+$ boson in the boosted-Higgs regime at N$^3$LO in QCD, fully differential in the kinematics of the final state. This calculation marks an important milestone in the ongoing effort to obtain N$^3$LO predictions for key processes at the LHC~\cite{Anastasiou:2015vya,Anastasiou:2016cez,Dreyer:2016oyx,Mistlberger:2018etf,Dreyer:2018qbw,Cieri:2018oms,Chen:2019lzz,Chen:2019fhs,Duhr:2019kwi,Duhr:2020sdp,Duhr:2021vwj,Chen:2021isd,Billis:2021ecs,Camarda:2021ict,Chen:2021vtu,Chen:2022cgv,Baglio:2022wzu,Neumann:2022lft,Chen:2022lwc,Campbell:2023lcy,Czakon:2026zhi,Chen:2026zmi}. We present state-of-the-art QCD predictions for the cross section and the Higgs boson transverse-momentum distribution considering fiducial selection cuts relevant for current experimental analyses at the LHC. 
The perturbative series for this process in the boosted regime exhibits poor convergence through NNLO, making the N$^3$LO prediction necessary for reliable phenomenology.
Our results reach percent-level QCD precision, reducing the residual dependence on perturbative scales, and significantly improving the viability of Higgs-strahlung as a precision probe of the Standard Model.
 
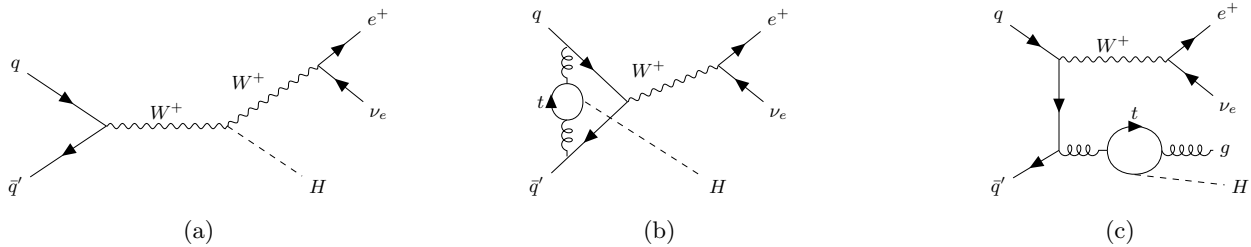
\begin{figure*}[t]
\centering
\begin{subfigure}[b]{0.32\textwidth}
\centering
\scalebox{0.8}{
\begin{tikzpicture}
\begin{feynman}
  \vertex (q1) at (-2.5, 1)  {\(q\)};
  \vertex (q2) at (-2.5,-1)  {\(\bar{q}'\)};
  \vertex (v1) at (-1, 0);
  \vertex (v2) at ( 1, 0);
  \vertex (v3) at ( 2.5, 1);
  \vertex (lp) at ( 3.5, 1.8) {\(e^+\)};
  \vertex (nu) at ( 3.5, 0.2) {\(\nu_e\)};
  \vertex (H)  at ( 2.5,-1)  {\(H\)};
  \diagram*{
    (q1) -- [fermion] (v1) -- [fermion] (q2),
    (v1) -- [boson, edge label=\(W^+\)] (v2),
    (v2) -- [boson, edge label=\(W^+\)] (v3),
    (v3) -- [fermion] (lp),
    (nu) -- [fermion] (v3),
    (v2) -- [scalar] (H),
  };
\end{feynman}
\end{tikzpicture}
}
\caption{\label{fig:Born}}
\end{subfigure}
\hfill
\begin{subfigure}[b]{0.32\textwidth}
\centering
\scalebox{0.8}{
\begin{tikzpicture}
\begin{feynman}
  \vertex (q1) at (-2.5, 1.4)  {\(q\)};
  \vertex (q1p) at (-2.0, 0.9);
  \vertex (q2) at (-2.5,-1.4)  {\(\bar{q}'\)};
  \vertex (q2p) at (-2.0, -0.9);
  \vertex (loop) at (-1.7, 0);
  \vertex (h) at (0.5, -1.4) {\(H\)};
  \vertex (v1) at (-1, 0);
  \vertex (v2) at (0.5, 0.6);
  \vertex (lp) at ( 1.5, 1.4) {\(e^+\)};
  \vertex (nu) at ( 1.5, -0.2) {\(\nu_e\)};
  \vertex (t1) at (-2, 0.3);
  \vertex (t2) at (-2, -0.3);
  \diagram*{
    (q1) -- [fermion] (v1),
    (v1) -- [fermion] (q2),
    (q1p) -- [gluon] (t1),
    (t2)  -- [gluon] (q2p),
    (t1) -- [half left] (t2),
    (t2) -- [fermion, half left, edge label=\(t\)] (t1),
    (loop) -- [scalar] (h),
    (v1) -- [boson, edge label=\(W^+\)] (v2),
    (nu) -- [fermion] (v2) -- [fermion] (lp),
  };
\end{feynman}
\end{tikzpicture}
}
\caption{\label{fig:ytop1}}
\end{subfigure}
\hfill
\begin{subfigure}[b]{0.32\textwidth}
\centering
\scalebox{0.8}{
\begin{tikzpicture}
\begin{feynman}
  \vertex (q1) at (-2.5, 1.8)  {\(q\)};
  \vertex (q1p) at (-1.5, 1.1);
  \vertex (q2p) at (-1.5, -0.4);
  \vertex (q2) at (-2.5,-1)  {\(\bar{q}'\)};
  \vertex (j1) at (-0.70, -0.4);
  \vertex (j2) at ( 0.20, -0.4);
  \vertex (loop) at (-0.25, -0.8 );
  \vertex (H) at (1.5, -1) {\(H\)};
  \vertex (j3) at ( 1.25, -0.4) {\(g\)};
  \vertex (v1) at (0.3, 1.1);
  \vertex (lp) at ( 1.3, 1.9) {\(e^+\)};
  \vertex (nu) at ( 1.3, 0.3) {\(\nu_e\)};
  \diagram*{
    (q1) -- [fermion] (q1p) -- [fermion] (q2p) -- [fermion] (q2),
    (q1p) -- [boson, edge label=\(W^+\)] (v1),
    (nu) -- [fermion] (v1) -- [fermion] (lp),
    (q2p) -- [gluon] (j1),
    (j1) -- [half right] (j2) -- [gluon] (j3),
    (j1) -- [fermion, half left, edge label = \(t\)] (j2),
    (loop) -- [scalar] (H),
  };
\end{feynman}
\end{tikzpicture}
}
\caption{\label{fig:ytop2}}
\end{subfigure}
\caption{(a) Born-level Feynman diagram for the $pp \rightarrow W^+(\rightarrow e^+\nu_e)H$ Higgs-strahlung process. (b), (c) Representative Feynman diagrams for top-quark Yukawa contributions including (b) purely virtual corrections and (c) an additional real emission.}
\end{figure*}

\paragraph*{Methodology. ---}
We consider the Higgs-strahlung process \mbox{$pp \rightarrow W^+(\rightarrow e^+\nu_e)H$}, depicted at Born-level in Fig.~\ref{fig:Born}.
The calculation presented in this Letter combines fixed-order perturbative predictions with elements of transverse-momentum resummation within the framework of $q_T$-slicing~\cite{Catani:2007vq}, that has been recently employed to obtain an array of state-of-the-art N$^3$LO predictions in other key color-singlet production processes~\cite{Cieri:2018oms,Billis:2021ecs,Camarda:2021ict,Chen:2021vtu,Chen:2022cgv,Neumann:2022lft,Chen:2022lwc,Campbell:2023lcy,Czakon:2026zhi,Chen:2026zmi}.
This approach is based on dissecting the radiation phase space using a \emph{slicing cut} ($\qtcut$) on the transverse momentum $q_T$ of the color-singlet system (in our case $e^+\nu_e H$, which we dub $WH$ henceforth). The fully-differential cross section is decomposed, schematically, as
\begin{equation}\label{eq:slicing}
\sigma = \int_{q_T < \qtcut} \! \mathrm{d}\sigma^{\rm singular} + \int_{q_T > \qtcut} \! \mathrm{d}\sigma^{\rm regular}.
\end{equation}
The contribution below the cut, $\mathrm{d}\sigma^{\rm singular}$, is computed using the N$^3$LO expansion of the leading-power resummation formula of the transverse-momentum spectrum,
for which all ingredients at this perturbative order have been obtained in the literature~\cite{Gehrmann:2010ue,Catani:2012qa,Gehrmann:2014yya,Lubbert:2016rku,Echevarria:2016scs,Li:2016ctv,Vladimirov:2016dll,Luo:2019szz,Ebert:2020yqt,Luo:2020epw}.
In this Letter, $\mathrm{d}\sigma^{\rm singular}$ is computed with the \texttt{RadISH} framework~\cite{Monni:2016ktx, Bizon:2017rah, Re:2021con} for transverse-momentum resummation.
The above-cut contribution $\mathrm{d}\sigma^{\rm regular}$ is evaluated with an NNLO calculation for the \mbox{$pp\to W^+H+$1 jet} process, obtained using the \texttt{NNLOJET} framework~\cite{NNLOJET:2025rno}, which computes higher-order QCD corrections with the antenna subtraction method~\cite{Gehrmann-DeRidder:2005btv,Currie:2013vh}. The implementation of the \mbox{$pp\to W^+H+$1 jet} process in \texttt{NNLOJET} is described in detail in Refs.~\cite{Gauld:2020ced,Gauld:2021ule} and relies on \texttt{OpenLoops2}~\cite{Buccioni:2019sur} for the evaluation of necessary scattering amplitudes at NNLO.
Owing to the approximations inherent in the computation of the below-cut contribution, the slicing method yields predictions that are accurate up to power corrections in $\qtcut/Q$, where $Q$ is the invariant mass of the color-singlet system. This requires the use of a $\qtcut$ value that is sufficiently small to suppress power corrections, while remaining large enough to ensure numerical stability in the above-cut calculation. 
We make use of the slicing procedure outlined above exclusively for the calculation of the N$^3$LO (${\cal O}(\alpha_s^3)$) correction. Instead, we rely on \texttt{NNLOJET}~\cite{Gauld:2019yng} for the fully-differential cross section up to NNLO.

At NNLO, alongside QCD corrections to the genuine Higgs-strahlung process, contributions where a Higgs boson couples to a top-quark loop also appear. These start at $\mathcal{O}(y_t\alpha_s^2$), with $y_t$ being the top-quark Yukawa coupling, and are numerically relevant for our precision target. The necessary amplitudes were calculated in Ref.~\cite{Brein:2011vx}.
The purely virtual corrections, depicted in Fig.~\ref{fig:ytop1}, are obtained in the heavy-top-quark limit, while the full top-mass dependence is retained for all contributions involving an additional real emission, depicted in Fig.~\ref{fig:ytop2}.
These amplitudes constitute a gauge-invariant set on their own and are both ultraviolet- and infrared-finite, therefore no subtraction procedure is needed for their evaluation. For these reasons, the $\mathcal{O}(y_t\alpha_s^2)$ corrections can be computed independently from the genuine $\mathcal{O}(\alpha_s^k)$ contributions and directly added at the cross-section level.
Corrections at higher orders in $y_t$ or $\alpha_s$, which are largely unknown, are neglected.
\begin{table*}[!]
\renewcommand{\arraystretch}{1.5}
\setlength{\tabcolsep}{20pt}
\begin{tabular}{ccccc}
  \toprule
  $\sigma\; [\pb]$ &
   Boosted &
   Boosted &
   STXS
  \\[1ex]
   &
  \multicolumn{1}{c}{ $\pth \geq 250$ GeV } &
  \multicolumn{1}{c}{ $\pth \geq 400$ GeV } &
  \multicolumn{1}{c}{ $\pth \geq 400$ GeV, $\ptw \geq 250$ GeV  } 
  \\
  \cmidrule(r{.5em}){1-1} 
  \cmidrule(r{.5em}){2-2} 
  \cmidrule(r{.5em}){3-3}
  \cmidrule(r{.5em}){4-4}
  LO &
  $2.457^{+2.5\%}_{-2.5\%}$ &
  $0.4639^{+4.7\%}_{-4.4\%}$ &
  $0.4639^{+4.7\%}_{-4.4\%}$ 
  \\
  NLO &
  $3.324^{+3.3\%}_{-2.8\%}$ &
  $0.6368^{+4.0\%}_{-3.5\%}$ &
  $0.5858^{+2.8\%}_{-2.5\%}$
  \\
  NNLO &
  $3.564^{+1.4\%}_{-1.4\%}$ &
  $0.6835^{+1.4\%}_{-1.7\%}$ &
  $0.6116^{+0.7\%}_{-1.0\%}$ 
  \\
  N$^3$LO &
  $(3.62\pm 0.01)^{+0.5\%}_{-0.8\%}$&
  $(0.694\pm 0.002)^{+0.7\%}_{-0.7\%}$ &
  $(0.621\pm 0.002)^{+0.6\%}_{-0.5\%}$ 
  \\
  \bottomrule
\end{tabular}
\caption{\label{tab:xs} Fiducial cross sections for the associated production a $W^+$ boson and a Higgs boson at the LHC at $\sqrt{s}=13.6$ TeV. The fiducial cuts are indicated in the header.
At N$^3$LO, results obtained with $\qtcut=5$ GeV are considered.
All quoted cross-section values come with the associated scale uncertainties, obtained as detailed in the text, which are reported as percentages of the respective central values.
For the N$^3$LO cross sections, the errors reported in brackets represent a conservative estimate of the numerical-integration uncertainty, including both the Monte Carlo error and the $q_T$-slicing error.}
\end{table*}

\paragraph*{Computational Setup ---}
We now discuss the setup used to produce the predictions presented in this Letter.
We consider proton-proton collisions at the LHC with a center-of-mass energy of $\sqrt{s} = 13.6$~TeV. The parton densities are described by the approximate N$^3$LO set \texttt{NNPDF40\_an3lo\_as\_01180}~\cite{NNPDF:2024nan} with \mbox{$\alpha_s(m_Z) = 0.118$}, accessed through the \texttt{LHAPDF} interface~\cite{Buckley:2014ana}. We use the same set at all perturbative orders.
The central factorization and renormalization scales $\mu_F$ and $\mu_R$ are set equal to the invariant mass $Q$ of the $WH$ system.
Missing higher-order uncertainties are estimated through canonical 7-point scale variations, varying $\mu_R$ and $\mu_F$ by a factor of two around their central value subject to the constraint \mbox{$1/4 < \mu_R/\mu_F < 4$}.
The relevant Standard Model parameters are set following the Run~III Higgs Cross Section Working Group recommendation~\cite{LHCHXSWG-RUNIII}.
In particular, the Higgs, $W$, $Z$, and top-quark masses are set to \mbox{$m_H = 125.09$~GeV}, \mbox{$m_W = 80.379$~GeV}, \mbox{$m_Z = 91.1876$~GeV}, and \mbox{$m_t = 172.5$~GeV}, while the $W$ width is \mbox{$\Gamma_W = 2.085$~GeV}.
The electroweak couplings are computed in the $G_\mu$ scheme with \mbox{$G_F = 1.1663788\times 10^{-5}$~GeV$^{-2}$}.

The focus of this Letter is the description of Higgs production in the boosted regime, defined by requiring the Higgs boson to have a transverse momentum $\pth$ above a given threshold.
We consider two setups in our phenomenological study.
The first, simply referred to as \emph{boosted}, is inclusive over the kinematics of the $W$-boson decay products. We implement two different thresholds for the Higgs, namely $\pth\geq 250$ GeV and $\pth\geq 400$ GeV.
The second setup is inspired by the simplified template cross section (STXS) analyses~\cite{Andersen:2016qtm,  Berger:2019wnu, ATLAS:2019yhn}. In addition to a $\pth$ cut, a requirement $\ptw \geq 250$~GeV is introduced on the $W$-boson transverse momentum; we refer to this as the \emph{STXS} setup.
In either setup, we study both the fiducial cross section and the $\pth$ distribution.

\paragraph*{Validation ---}
We first test our methodology by computing the inclusive cross section for $pp \rightarrow W^+H$, with an on-shell $W^+$, which was previously obtained in Ref.~\cite{Baglio:2022wzu}. We find excellent agreement between our prediction and the results in the literature, separately for each of the initial-state flavor channels. 
This check, discussed in detail in the supplemental material~\cite{supplemental},  provides a highly non-trivial confirmation of the validity of our methodology. Importantly, it allows us to estimate the systematic uncertainties related to the slicing procedure, anticipated above, and to identify an optimal range of $\qtcut$ values.
The validation study~\cite{supplemental} indicates that $\qtcut=5$ GeV provides a suitable reference value for the computation of the N$^3$LO correction.

Missing power corrections from the slicing procedure are generally of ${\cal O}((\qtcut/Q)^2)$, hence negligible at the typical values of the invariant mass of the $WH$ system, $Q \simeq 200$ GeV. This is the case for the validation at the inclusive level, as well as for the boosted and STXS fiducial setups defined above, provided that asymmetric cuts are applied on $\pth$ and $\ptw$. These are sometimes referred to as \emph{staggered} cuts and are possible in this case because the decay products of the $W^+$ boson and the Higgs boson are distinguishable.

\begin{figure}
\includegraphics[width = 0.5\textwidth]{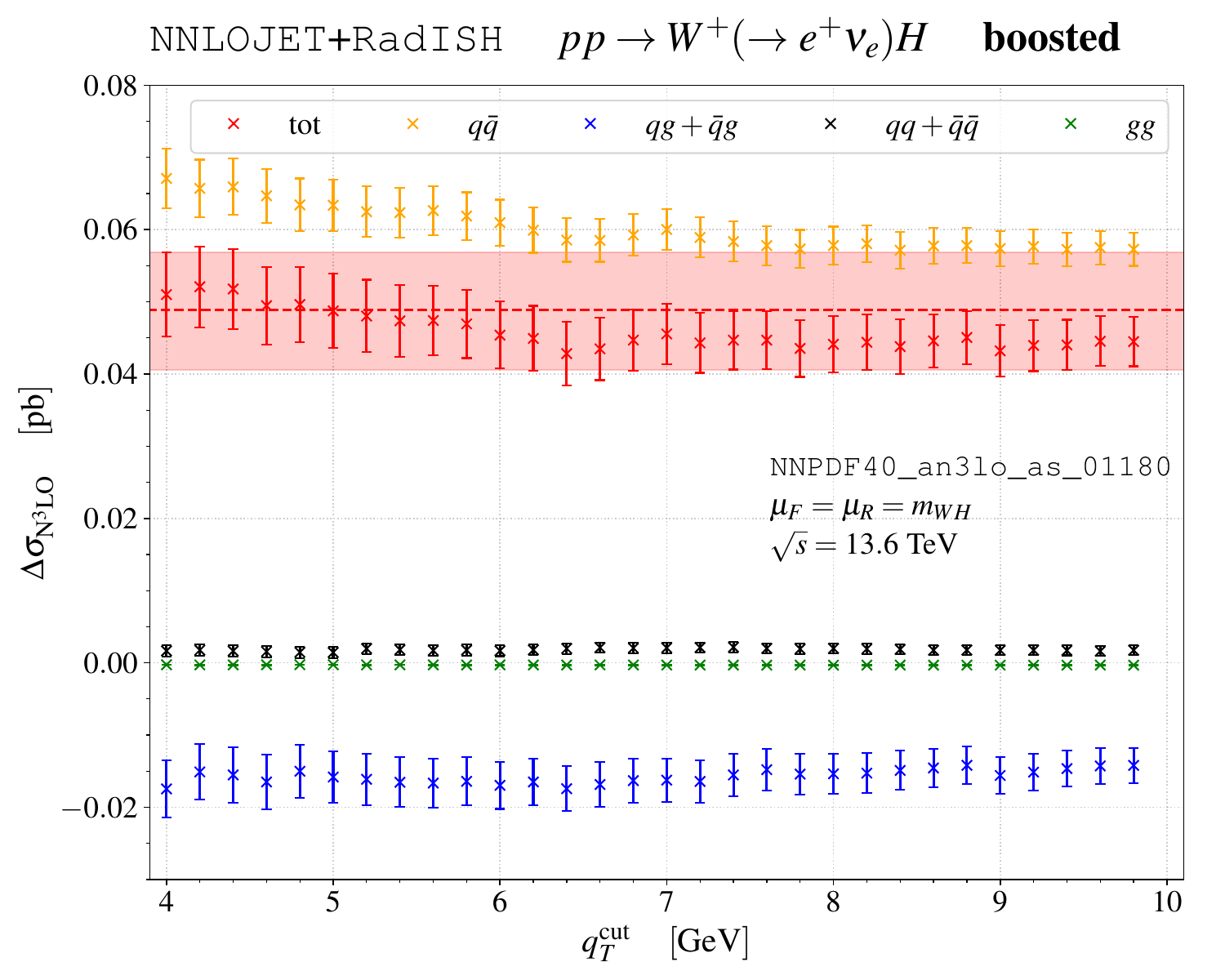}
\caption{N$^3$LO correction to the fiducial
cross section for the associated $W^+H$ production process at the LHC, with $\pth\geq 250$ GeV. The results are decomposed into partonic channels: $q\bar{q}$ (yellow), $qg+\bar{q}g$ (blue), $qq+\bar{q}\bar{q}$ (black), $gg$ (green) and total (red). Error bars indicate the Monte Carlo integration uncertainty. The horizontal band represents our best determination of the N$^3$LO correction, with the associated slicing error.}
\label{fig:cumulant_boosted}
\end{figure}

In Fig.~\ref{fig:cumulant_boosted}, we display the $\qtcut$ dependence of the fiducial N$^3$LO correction in the boosted setup, for each initial-state flavor channel separately. We note that the N$^3$LO correction receives significant contributions from the
$q\bar{q}$ and $qg+\bar{q}g$ channels, which partially cancel each other.

The remarkable stability of the result, which can be traced to the absence of linear power corrections of ${\cal O}(\qtcut/Q)$, allows us to opt for a default choice of $\qtcut=5$ GeV, as supported also by the validation of the method in Ref.~\cite{supplemental}.
We estimate a $q_T$-slicing error by varying the adopted $\qtcut$ value by $\pm1$ GeV with respect to the central choice, and taking the envelope of this variation including the Monte Carlo statistical error on the associated N$^3$LO correction. This is displayed by the band in Fig.~\ref{fig:cumulant_boosted}, which shows that this procedure provides a sufficiently conservative estimate of the slicing error.

The runtime of the N$^3$LO computation is dominated by the above-cut \mbox{$pp\to W^+H+1$ jet} component, and in particular its double-real contribution. The total computing time amounts to about 3M core hours, of which 2.5M are taken by the dominant $q\bar{q}$ channel, followed by $qg+\bar{q}g$ (0.4M core hours).

\paragraph*{Results. ---}
Fixed-order results for the fiducial cross sections from LO to N$^3$LO for both setups are collected in Table~\ref{tab:xs}.
All cross-section values appearing in the table display theoretical uncertainties stemming from the 7-point scale variation discussed earlier, which are reported as percentages of the respective central values. The
N$^3$LO cross sections feature an extra (absolute) error, shown in brackets, encoding the estimate of the $q_T$-slicing error, obtained as detailed above.

In the boosted setup, for both values of the $\pth$ cut (first and second column of Tab.~\ref{tab:xs}), the perturbative series up to NNLO exhibits poor convergence, as assessed through scale-variation bands.
Neither the NLO nor the NNLO value lie within the uncertainty band of the preceding order, and only at N$^3$LO does the series begin to show signs of convergence.
The N$^3$LO correction amounts to approximately $+1.5\%$, and is comparable in size to the scale uncertainty of the NNLO result, while the residual scale-variation band at N$^3$LO falls below the percent level. 
We stress that the N$^3$LO correction in the boosted regime cannot simply be obtained by multiplying the inclusive result by a phase-space acceptance factor: indeed the inclusive N$^3$LO correction is negative ($-1\%$)~\cite{Baglio:2022wzu}, with no reduction in the scale uncertainty with respect to NNLO. This aspect further highlights the importance of a fully differential control over the phase space for precision Higgs phenomenology.
We note that the N$^3$LO correction in the boosted regime is comparable in size with the top-Yukawa-induced contributions at order $\alpha_s^2$, which amount to $+1\%$ of the total cross section.

In the STXS setup (third column of Tab.~\ref{tab:xs}), defined by $\pth \geq 400$ GeV and $\ptw \geq 250$ GeV, perturbative corrections are smaller in magnitude with respect to the boosted case. Scale uncertainty, which appears to largely underestimate the true size of missing higher-order corrections, is already below the percent level at NNLO.
Perturbative convergence is moderately improved upon inclusion of the N$^3$LO correction, which amounts to roughly $+1.5$\%, and lies just outside of the NNLO scale-variation band. The top-Yukawa contributions induce approximately a $0.5\%$ increase in the total cross section, and their magnitude is comparable with the residual numerical error of the computation.

\begin{figure*}[t]
\includegraphics[width = 0.48\textwidth]{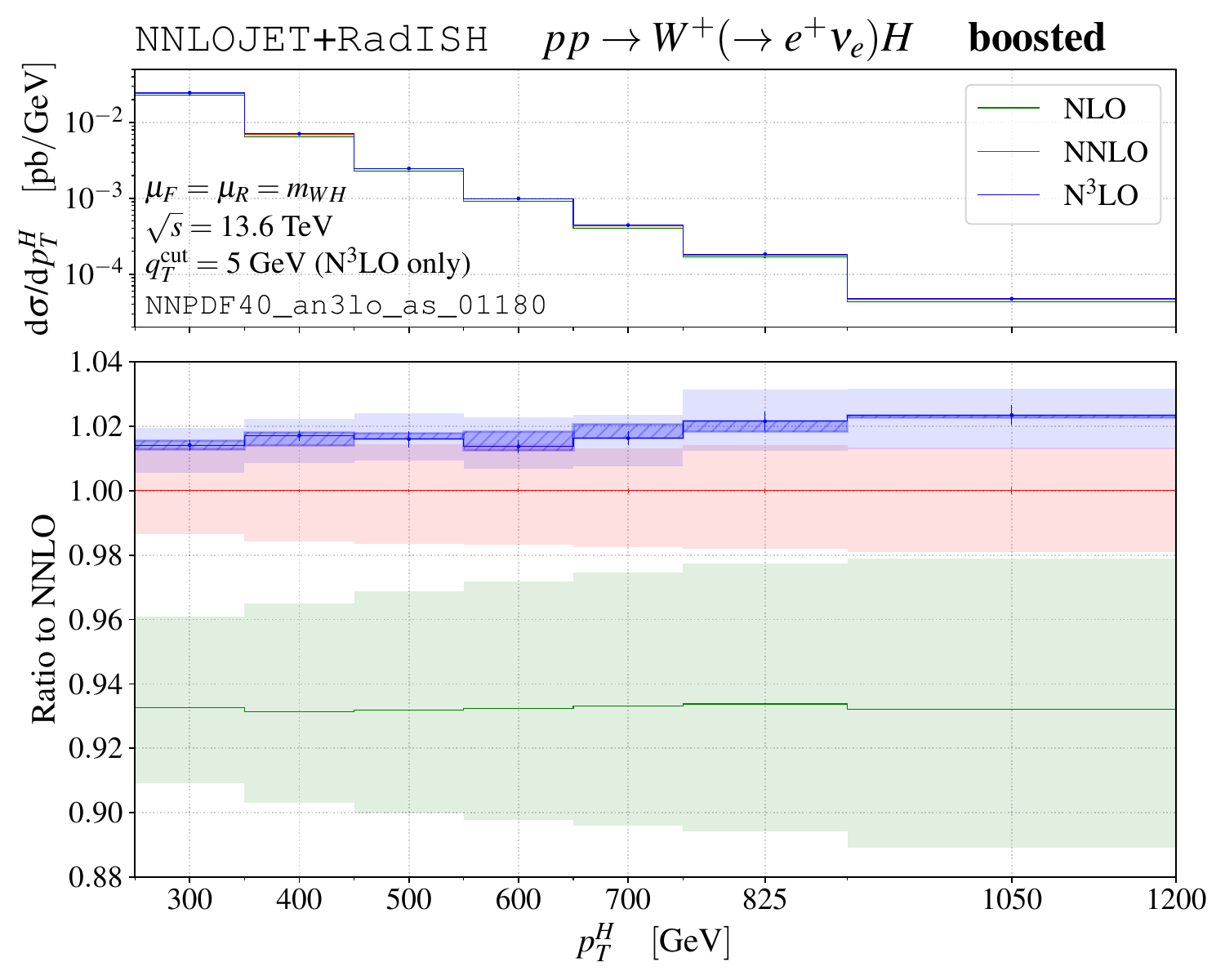}
\includegraphics[width = 0.48\textwidth]{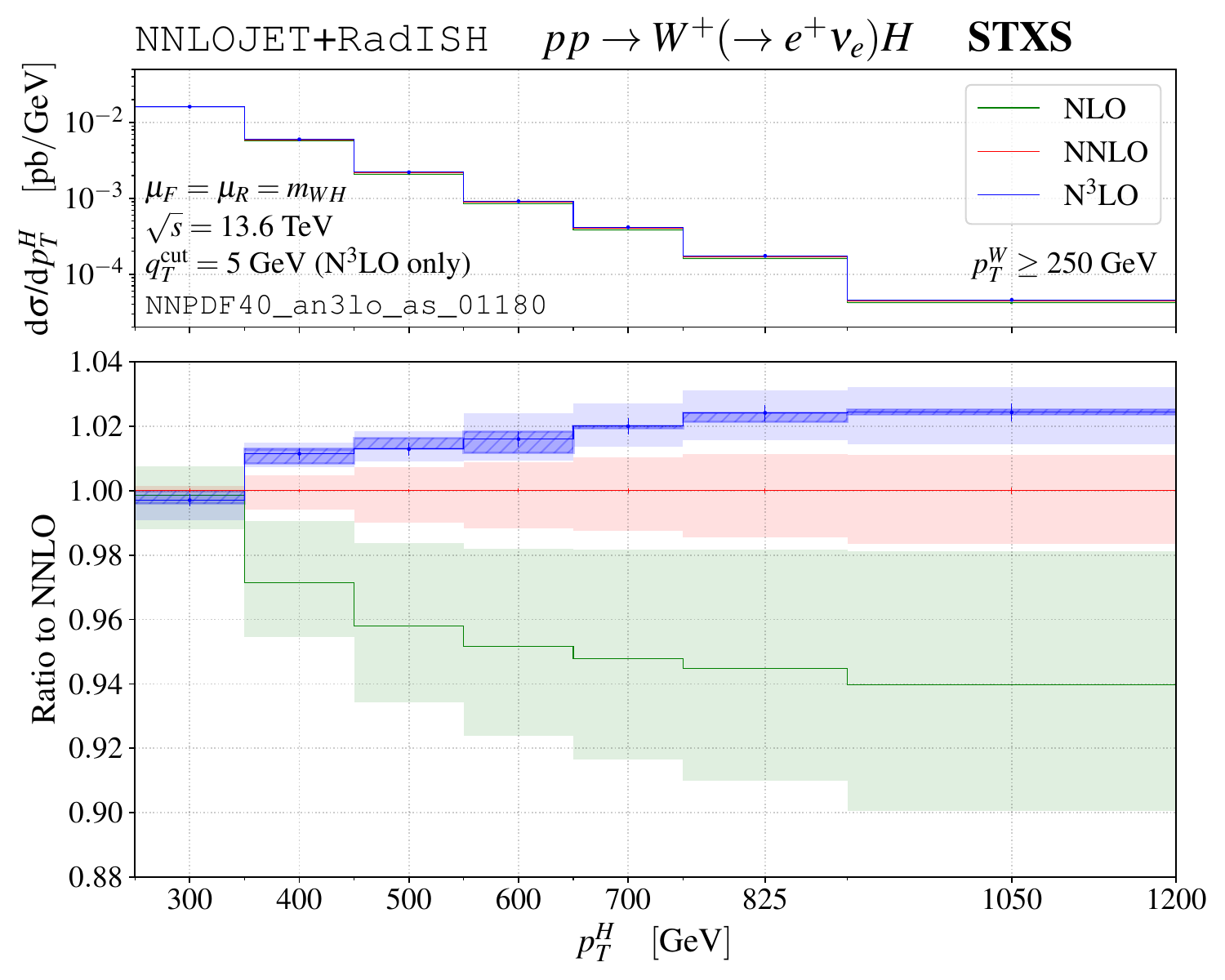}
\caption{Fiducial $\pth$ distribution at NLO (green), NNLO (red) and N$^3$LO (blue) in the associated $W^+H$ production process, with no cuts on $\ptw$ (boosted, left) and with $\ptw\geq250$ GeV (STXS, right). Lighter shaded bands represent scale-variation uncertainties, while the darker blue bands at N$^3$LO are obtained by varying the slicing parameter $\qtcut$ between 4 and 6 GeV.}
\label{fig:pth}
\end{figure*}

Our framework also allows us to present, for the first time, N$^3$LO-accurate differential distributions for the considered process.
In Fig.~\ref{fig:pth}, we show the fiducial $\pth$ spectrum at NLO, NNLO and N$^3$LO for the boosted (left panel) and STXS (right panel) setups in the range \mbox{$250 \leq \pth \leq 1200$~GeV}.
In the boosted case, the N$^3$LO correction amounts to an increase over the
NNLO result between $1.5\%$ and $2\%$, relatively stable across the whole $\pth$ range considered. 
The residual missing higher-order uncertainty estimated via scale variations (lighter shaded bands in Fig.~\ref{fig:pth}) is at the $1\%$ level at $\pth \simeq 300$~GeV, growing to $2\%$ in the tail of the distribution.
The uncertainty associated with the $q_T$-slicing procedure (darker shaded bands in Fig.~\ref{fig:pth}) remains well below the scale-variation band throughout the displayed range, highlighting the excellent numerical control of our differential predictions.

In the STXS setup, the N$^3$LO contribution instead induces a more significant shape distortion, with a small negative correction, below the percent level, near \mbox{$\pth \simeq 250$}~GeV, and a larger positive correction,
at the $2$--$3\%$ level, in the tail. In the large-$\pth$ limit, the perturbative progression is similar to the boosted case, as the sensitivity to the $\ptw$ cut diminishes.
Throughout most of the spectrum, the $q_T$-slicing uncertainty is smaller than the scale uncertainty, becoming comparable to the latter only at low $\pth$, owing to the reduced size of the missing-higher-order uncertainty.
The leftmost bin of the distribution is characterized by a symmetric-cut configuration $\pth,\,\ptw\geq250$ GeV, which is known to induce potentially large linear power corrections of ${\cal O}(\qtcut/Q)$~\cite{Frixione:1996ym,Grazzini:2017mhc,Ebert:2019zkb,Salam:2021tbm,Alekhin:2021xcu}. In our framework, we systematically calculate the linear power corrections using the approach of Refs.~\cite{Buonocore:2021tke,Camarda:2021jsw}, and add them to the below-cut contribution in Eq.~\eqref{eq:slicing}, thereby always reducing the residual $\qtcut$ dependence to quadratic order.

We further note that fiducial phase spaces characterized by symmetric cuts may suffer from poor numerical stability of the perturbative series, as pointed out  in Refs.~\cite{Klasen:1995xe,Harris:1997hz,Frixione:1997ks}.
Therefore, one may question the reliability of fixed-order predictions in the first bin, which contains configurations where the $W$ and the $H$ bosons are back-to-back.
A correct treatment of soft-radiation enhancements characterizing such configurations requires all-order resummation of potentially large logarithms of the imbalance. 
Although a precise answer to this question would demand a detailed study, which we do not pursue in this work, let us note that the large bin width of 100 GeV allows the bin to be populated by staggered configurations on top of back-to-back ones, thus mitigating the sensitivity to soft radiation.

\paragraph*{Conclusions and Outlook. ---}
In this Letter, we have presented the first fully differential N$^3$LO QCD prediction for the Higgs-strahlung process \mbox{$pp\to W^+(\to e^+\nu_e)H$} in the boosted-Higgs regime at the LHC, defined by a lower cut on the Higgs transverse momentum. Within the $q_T$-slicing framework, our calculation combines the N$^3$LO expansion of transverse-momentum resummation with NNLO predictions for \mbox{$pp\to W^+(\to e^+\nu_e)H$ + 1 jet} production, and includes the numerically relevant top-Yukawa-induced contributions at ${\cal O}(y_t\alpha_s^2)$. We have obtained state-of-the-art predictions for fiducial cross sections and Higgs transverse-momentum distributions, relevant for current and future boosted-Higgs analyses at the LHC and HL-LHC.

We find that the N$^3$LO corrections increase the NNLO prediction by approximately $1-2\%$ in the boosted setup, while significantly reducing the residual perturbative uncertainty to the percent level or below. The dependence of the N$^3$LO corrections on the Higgs transverse momentum cut strongly varies with the fiducial setup under consideration, and in particular with the cuts on the decay products of the accompanying $W^+$ boson. This highlights the importance of fully-differential computations for an accurate description of the boosted Higgs kinematics.

Our results demonstrate that perturbative convergence, which is rather poor up to NNLO in QCD, is improved upon inclusion of the N$^3$LO corrections, highlighting the importance of this perturbative order for precision Higgs-strahlung phenomenology. These predictions constitute an important step towards matching the precision expected from the HL-LHC Higgs program and strengthen the role of boosted Higgs production as a sensitive probe of Standard Model dynamics and possible new-physics effects.

The techniques used in this work can also be applied to $ZH$ associate production.
In this case, however, loop-induced contributions to $gg \to ZH$, entering at ${\cal O}(\as^2)$, should also be considered, as they are of crucial phenomenological importance due to the large gluon flux at the LHC.
These contributions can be included separately, as they are gauge invariant and do not interfere with the corrections stemming from the topology in Fig~\ref{fig:Born}.

The level of precision reached in this work challenges that of theoretical inputs beyond perturbative QCD, such as PDFs and $\as$ \cite{Karlberg:2024zxx}. Moreover, it is well known that EW corrections have a large effect on the Higgs-strahlung cross section at large $\pth$~\cite{Granata:2017iod}.
Therefore, further investigation of both pure EW and mixed QCD+EW effects is required to reach the percent-level target for Higgs precision phenomenology at the LHC.

\begin{acknowledgments}
  We wish to thank Thomas Gehrmann and Rhorry Gauld for discussions and comments on the article. MM is supported by a Royal Society Newton International Fellowship (NIF/R1/232539).
  AG is supported by the Swiss National Science Foundation (SNF) under contract 200021-231259.
  The work of PM is funded by the European Union (ERC, grant agreement
  No. 101044599, JANUS).
  Views and opinions expressed are however those of the authors only
  and do not necessarily reflect those of the European Union or the
  European Research Council Executive Agency. Neither the European
  Union nor the granting authority can be held responsible for them.
  FS is supported by the STFC under grant agreement ST/P006744/1.
\end{acknowledgments}

\bibliographystyle{apsrev4-2}
\bibliography{references}

\input{supplemental_material}
\end{document}

%% file: supplemental_material.tex
\newpage

\onecolumngrid
\newpage
\appendix

%

\section{Supplementary material}
In this Appendix we validate our implementation by computing the inclusive cross section for $pp \rightarrow W^+H$ and comparing it against results obtained with the \texttt{n3loxs} code~\cite{Baglio:2022wzu}. The kinematical setup follows that of the calculation presented in the Letter, but no transverse-momentum cut is imposed on the Higgs or $W^+$ boson. The $W^+$ boson is taken to be on-shell, and contributions proportional to the top-quark Yukawa coupling, which are absent in \texttt{n3loxs}, are omitted.

\begin{figure}[t]
\includegraphics[width = 0.5\textwidth]{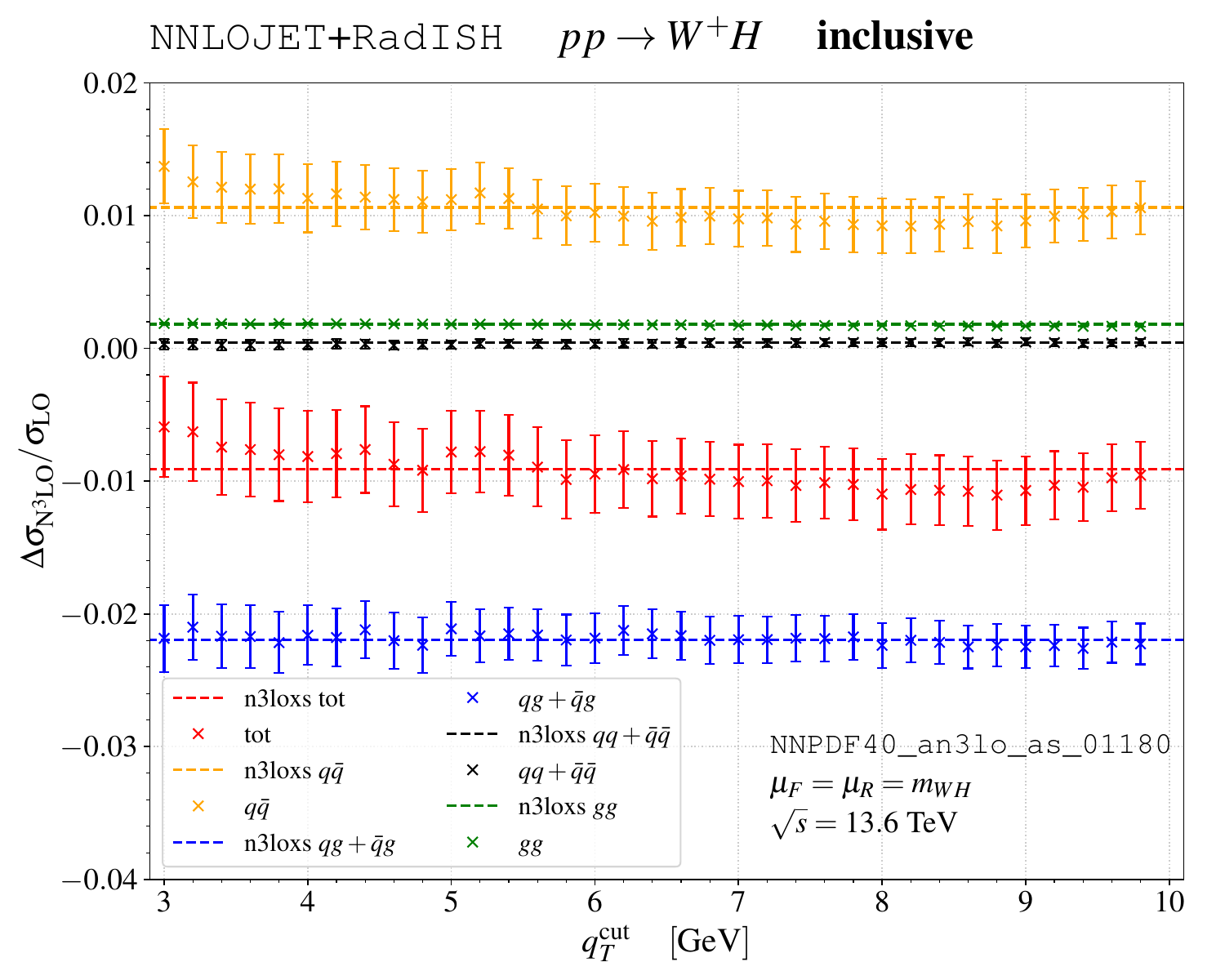}
\caption{N$^3$LO correction to the inclusive
cross section for the associated $W^+H$ production process at the LHC, normalized to the LO cross section. The results obtained in this Letter with the $q_T$-slicing method as a function of $\qtcut$ (crosses) are compared with reference numbers (dashed lines) obtained with the \texttt{n3loxs} code~\cite{Baglio:2022wzu}. The results are decomposed into partonic channels: $q\bar{q}$ (yellow), $qg+\bar{q}g$ (blue), $qq+\bar{q}\bar{q}$ (black), $gg$ (green) and total (red). Error bars indicate the Monte Carlo integration uncertainty. }
\label{fig:inclusivexs}
\end{figure}

Up to NNLO, we find perfect agreement between \texttt{n3loxs} and the results obtained with \texttt{NNLOJET} using the antenna subtraction method. As described in the Letter, the N$^3$LO correction is computed with the $q_T$-slicing technique. In Fig.~\ref{fig:inclusivexs}, we show the N$^3$LO correction to the inclusive cross section as a function of $\qtcut$, decomposed into initial-state flavor channels and compared with the numerical results of \texttt{n3loxs}. We find excellent agreement across the full range $3\text{ GeV}\leq\qtcut\leq 10\text{ GeV}$, with residual variations well within the Monte Carlo integration uncertainty of the N$^3$LO correction. The latter reaches at most $\pm 30\%$ in the $q\bar{q}$ channel.
The $q\bar{q}$ and $qg+\bar{q}g$ channels make the dominant contribution. However, in the total correction, the former partially cancels against the latter, leading to a larger relative uncertainty. Conversely, the $qq+\bar{q}\bar{q}$, and $gg$ components are numerically small and phenomenologically negligible. Towards $\qtcut\sim 3$ GeV, the $q\bar{q}$ contribution shows a mild systematic deviation from the \texttt{n3loxs} result, indicating potential numerical instabilities that inevitably arise in the NNLO calculation of $pp\to W^+H$ + 1 jet as $\qtcut\to 0$.

Based on the above analysis, we adopt $\qtcut = 5$ GeV as the reference slicing cut for the results presented in the Letter. This choice minimizes the systematic uncertainty stemming from slicing power corrections, while remaining safely above the region where numerical instabilities may develop (around $\qtcut \sim 3$ GeV). We also estimate the systematic uncertainty associated with the choice of $\qtcut$ by varying the latter by $\pm 1$ GeV around the reference value. The uncertainty is obtained by taking the envelope of the shift in the value of the cross section and the associated Monte Carlo error.





%% file: main.bbl
\begin{thebibliography}{89}%
\makeatletter
\providecommand \@ifxundefined [1]{%
 \@ifx{#1\undefined}
}%
\providecommand \@ifnum [1]{%
 \ifnum #1\expandafter \@firstoftwo
 \else \expandafter \@secondoftwo
 \fi
}%
\providecommand \@ifx [1]{%
 \ifx #1\expandafter \@firstoftwo
 \else \expandafter \@secondoftwo
 \fi
}%
\providecommand \natexlab [1]{#1}%
\providecommand \enquote  [1]{``#1''}%
\providecommand \bibnamefont  [1]{#1}%
\providecommand \bibfnamefont [1]{#1}%
\providecommand \citenamefont [1]{#1}%
\providecommand \href@noop [0]{\@secondoftwo}%
\providecommand \href [0]{\begingroup \@sanitize@url \@href}%
\providecommand \@href[1]{\@@startlink{#1}\@@href}%
\providecommand \@@href[1]{\endgroup#1\@@endlink}%
\providecommand \@sanitize@url [0]{\catcode `\\12\catcode `\$12\catcode
  `\&12\catcode `\#12\catcode `\^12\catcode `\_12\catcode `\%12\relax}%
\providecommand \@@startlink[1]{}%
\providecommand \@@endlink[0]{}%
\providecommand \url  [0]{\begingroup\@sanitize@url \@url }%
\providecommand \@url [1]{\endgroup\@href {#1}{\urlprefix }}%
\providecommand \urlprefix  [0]{URL }%
\providecommand \Eprint [0]{\href }%
\providecommand \doibase [0]{https://doi.org/}%
\providecommand \selectlanguage [0]{\@gobble}%
\providecommand \bibinfo  [0]{\@secondoftwo}%
\providecommand \bibfield  [0]{\@secondoftwo}%
\providecommand \translation [1]{[#1]}%
\providecommand \BibitemOpen [0]{}%
\providecommand \bibitemStop [0]{}%
\providecommand \bibitemNoStop [0]{.\EOS\space}%
\providecommand \EOS [0]{\spacefactor3000\relax}%
\providecommand \BibitemShut  [1]{\csname bibitem#1\endcsname}%
\let\auto@bib@innerbib\@empty
\bibitem [{\citenamefont {Aad}\ \emph {et~al.}(2021)\citenamefont {Aad} \emph
  {et~al.}}]{ATLAS:2020fcp}%
  \BibitemOpen
  \bibfield  {author} {\bibinfo {author} {\bibfnamefont {G.}~\bibnamefont
  {Aad}} \emph {et~al.} (\bibinfo {collaboration} {ATLAS}),\ }\href
  {https://doi.org/10.1140/epjc/s10052-020-08677-2} {\bibfield  {journal}
  {\bibinfo  {journal} {Eur. Phys. J. C}\ }\textbf {\bibinfo {volume} {81}},\
  \bibinfo {pages} {178} (\bibinfo {year} {2021})},\ \Eprint
  {https://arxiv.org/abs/2007.02873} {arXiv:2007.02873 [hep-ex]} \BibitemShut
  {NoStop}%
\bibitem [{\citenamefont {Aad}\ \emph {et~al.}(2022)\citenamefont {Aad} \emph
  {et~al.}}]{ATLAS:2021tbi}%
  \BibitemOpen
  \bibfield  {author} {\bibinfo {author} {\bibfnamefont {G.}~\bibnamefont
  {Aad}} \emph {et~al.} (\bibinfo {collaboration} {ATLAS}),\ }\href
  {https://doi.org/10.1103/PhysRevD.105.092003} {\bibfield  {journal} {\bibinfo
   {journal} {Phys. Rev. D}\ }\textbf {\bibinfo {volume} {105}},\ \bibinfo
  {pages} {092003} (\bibinfo {year} {2022})},\ \Eprint
  {https://arxiv.org/abs/2111.08340} {arXiv:2111.08340 [hep-ex]} \BibitemShut
  {NoStop}%
\bibitem [{\citenamefont {Aad}\ \emph {et~al.}(2024)\citenamefont {Aad} \emph
  {et~al.}}]{ATLAS:2023jdk}%
  \BibitemOpen
  \bibfield  {author} {\bibinfo {author} {\bibfnamefont {G.}~\bibnamefont
  {Aad}} \emph {et~al.} (\bibinfo {collaboration} {ATLAS}),\ }\href
  {https://doi.org/10.1103/PhysRevLett.132.131802} {\bibfield  {journal}
  {\bibinfo  {journal} {Phys. Rev. Lett.}\ }\textbf {\bibinfo {volume} {132}},\
  \bibinfo {pages} {131802} (\bibinfo {year} {2024})},\ \Eprint
  {https://arxiv.org/abs/2312.07605} {arXiv:2312.07605 [hep-ex]} \BibitemShut
  {NoStop}%
\bibitem [{\citenamefont {Sirunyan}\ \emph {et~al.}(2020)\citenamefont
  {Sirunyan} \emph {et~al.}}]{CMS:2020zge}%
  \BibitemOpen
  \bibfield  {author} {\bibinfo {author} {\bibfnamefont {A.~M.}\ \bibnamefont
  {Sirunyan}} \emph {et~al.} (\bibinfo {collaboration} {CMS}),\ }\href
  {https://doi.org/10.1007/JHEP12(2020)085} {\bibfield  {journal} {\bibinfo
  {journal} {JHEP}\ }\textbf {\bibinfo {volume} {12}},\ \bibinfo {pages}
  {085}},\ \Eprint {https://arxiv.org/abs/2006.13251} {arXiv:2006.13251
  [hep-ex]} \BibitemShut {NoStop}%
\bibitem [{\citenamefont {Sirunyan}\ \emph {et~al.}(2021)\citenamefont
  {Sirunyan} \emph {et~al.}}]{CMS:2021nnc}%
  \BibitemOpen
  \bibfield  {author} {\bibinfo {author} {\bibfnamefont {A.~M.}\ \bibnamefont
  {Sirunyan}} \emph {et~al.} (\bibinfo {collaboration} {CMS}),\ }\href
  {https://doi.org/10.1103/PhysRevD.104.052004} {\bibfield  {journal} {\bibinfo
   {journal} {Phys. Rev. D}\ }\textbf {\bibinfo {volume} {104}},\ \bibinfo
  {pages} {052004} (\bibinfo {year} {2021})},\ \Eprint
  {https://arxiv.org/abs/2104.12152} {arXiv:2104.12152 [hep-ex]} \BibitemShut
  {NoStop}%
\bibitem [{\citenamefont {Hayrapetyan}\ \emph {et~al.}(2024)\citenamefont
  {Hayrapetyan} \emph {et~al.}}]{CMS:2024ddc}%
  \BibitemOpen
  \bibfield  {author} {\bibinfo {author} {\bibfnamefont {A.}~\bibnamefont
  {Hayrapetyan}} \emph {et~al.} (\bibinfo {collaboration} {CMS}),\ }\href
  {https://doi.org/10.1007/JHEP12(2024)035} {\bibfield  {journal} {\bibinfo
  {journal} {JHEP}\ }\textbf {\bibinfo {volume} {12}},\ \bibinfo {pages}
  {035}},\ \Eprint {https://arxiv.org/abs/2407.08012} {arXiv:2407.08012
  [hep-ex]} \BibitemShut {NoStop}%
\bibitem [{\citenamefont {Baglio}\ \emph {et~al.}(2020)\citenamefont {Baglio},
  \citenamefont {Dawson}, \citenamefont {Homiller}, \citenamefont {Lane},\ and\
  \citenamefont {Lewis}}]{Baglio:2020oqu}%
  \BibitemOpen
  \bibfield  {author} {\bibinfo {author} {\bibfnamefont {J.}~\bibnamefont
  {Baglio}}, \bibinfo {author} {\bibfnamefont {S.}~\bibnamefont {Dawson}},
  \bibinfo {author} {\bibfnamefont {S.}~\bibnamefont {Homiller}}, \bibinfo
  {author} {\bibfnamefont {S.~D.}\ \bibnamefont {Lane}},\ and\ \bibinfo
  {author} {\bibfnamefont {I.~M.}\ \bibnamefont {Lewis}},\ }\href
  {https://doi.org/10.1103/PhysRevD.101.115004} {\bibfield  {journal} {\bibinfo
   {journal} {Phys. Rev. D}\ }\textbf {\bibinfo {volume} {101}},\ \bibinfo
  {pages} {115004} (\bibinfo {year} {2020})},\ \Eprint
  {https://arxiv.org/abs/2003.07862} {arXiv:2003.07862 [hep-ph]} \BibitemShut
  {NoStop}%
\bibitem [{\citenamefont {Bizon}\ \emph {et~al.}(2017)\citenamefont {Bizon},
  \citenamefont {Gorbahn}, \citenamefont {Haisch},\ and\ \citenamefont
  {Zanderighi}}]{Bizon:2016wgr}%
  \BibitemOpen
  \bibfield  {author} {\bibinfo {author} {\bibfnamefont {W.}~\bibnamefont
  {Bizon}}, \bibinfo {author} {\bibfnamefont {M.}~\bibnamefont {Gorbahn}},
  \bibinfo {author} {\bibfnamefont {U.}~\bibnamefont {Haisch}},\ and\ \bibinfo
  {author} {\bibfnamefont {G.}~\bibnamefont {Zanderighi}},\ }\href
  {https://doi.org/10.1007/JHEP07(2017)083} {\bibfield  {journal} {\bibinfo
  {journal} {JHEP}\ }\textbf {\bibinfo {volume} {07}},\ \bibinfo {pages}
  {083}},\ \Eprint {https://arxiv.org/abs/1610.05771} {arXiv:1610.05771
  [hep-ph]} \BibitemShut {NoStop}%
\bibitem [{\citenamefont {Bizo{\'n}}\ \emph {et~al.}(2022)\citenamefont
  {Bizo{\'n}}, \citenamefont {Caola}, \citenamefont {Melnikov},\ and\
  \citenamefont {R{\"o}ntsch}}]{Bizon:2021rww}%
  \BibitemOpen
  \bibfield  {author} {\bibinfo {author} {\bibfnamefont {W.}~\bibnamefont
  {Bizo{\'n}}}, \bibinfo {author} {\bibfnamefont {F.}~\bibnamefont {Caola}},
  \bibinfo {author} {\bibfnamefont {K.}~\bibnamefont {Melnikov}},\ and\
  \bibinfo {author} {\bibfnamefont {R.}~\bibnamefont {R{\"o}ntsch}},\ }\href
  {https://doi.org/10.1103/PhysRevD.105.014023} {\bibfield  {journal} {\bibinfo
   {journal} {Phys. Rev. D}\ }\textbf {\bibinfo {volume} {105}},\ \bibinfo
  {pages} {014023} (\bibinfo {year} {2022})},\ \Eprint
  {https://arxiv.org/abs/2106.06328} {arXiv:2106.06328 [hep-ph]} \BibitemShut
  {NoStop}%
\bibitem [{\citenamefont {Gauld}\ \emph {et~al.}(2024)\citenamefont {Gauld},
  \citenamefont {Haisch},\ and\ \citenamefont {Schnell}}]{Gauld:2023gtb}%
  \BibitemOpen
  \bibfield  {author} {\bibinfo {author} {\bibfnamefont {R.}~\bibnamefont
  {Gauld}}, \bibinfo {author} {\bibfnamefont {U.}~\bibnamefont {Haisch}},\ and\
  \bibinfo {author} {\bibfnamefont {L.}~\bibnamefont {Schnell}},\ }\href
  {https://doi.org/10.1007/JHEP01(2024)192} {\bibfield  {journal} {\bibinfo
  {journal} {JHEP}\ }\textbf {\bibinfo {volume} {01}},\ \bibinfo {pages}
  {192}},\ \Eprint {https://arxiv.org/abs/2311.06107} {arXiv:2311.06107
  [hep-ph]} \BibitemShut {NoStop}%
\bibitem [{\citenamefont {Bonetti}\ \emph {et~al.}(2025)\citenamefont
  {Bonetti}, \citenamefont {Harlander}, \citenamefont {Korneev}, \citenamefont
  {Long}, \citenamefont {Melnikov}, \citenamefont {R{\"o}ntsch},\ and\
  \citenamefont {Tagliabue}}]{Bonetti:2025hnb}%
  \BibitemOpen
  \bibfield  {author} {\bibinfo {author} {\bibfnamefont {M.}~\bibnamefont
  {Bonetti}}, \bibinfo {author} {\bibfnamefont {R.~V.}\ \bibnamefont
  {Harlander}}, \bibinfo {author} {\bibfnamefont {D.}~\bibnamefont {Korneev}},
  \bibinfo {author} {\bibfnamefont {M.-M.}\ \bibnamefont {Long}}, \bibinfo
  {author} {\bibfnamefont {K.}~\bibnamefont {Melnikov}}, \bibinfo {author}
  {\bibfnamefont {R.}~\bibnamefont {R{\"o}ntsch}},\ and\ \bibinfo {author}
  {\bibfnamefont {D.~M.}\ \bibnamefont {Tagliabue}},\ }\href
  {https://doi.org/10.1103/ct5x-hl46} {\bibfield  {journal} {\bibinfo
  {journal} {Phys. Rev. D}\ }\textbf {\bibinfo {volume} {112}},\ \bibinfo
  {pages} {034033} (\bibinfo {year} {2025})},\ \Eprint
  {https://arxiv.org/abs/2502.12846} {arXiv:2502.12846 [hep-ph]} \BibitemShut
  {NoStop}%
\bibitem [{\citenamefont {Becker}\ \emph {et~al.}(2024)\citenamefont {Becker}
  \emph {et~al.}}]{Becker:2020rjp}%
  \BibitemOpen
  \bibfield  {author} {\bibinfo {author} {\bibfnamefont {K.}~\bibnamefont
  {Becker}} \emph {et~al.},\ }\href
  {https://doi.org/10.21468/SciPostPhysCore.7.1.001} {\bibfield  {journal}
  {\bibinfo  {journal} {SciPost Phys. Core}\ }\textbf {\bibinfo {volume} {7}},\
  \bibinfo {pages} {001} (\bibinfo {year} {2024})},\ \Eprint
  {https://arxiv.org/abs/2005.07762} {arXiv:2005.07762 [hep-ph]} \BibitemShut
  {NoStop}%
\bibitem [{\citenamefont {Baglio}\ \emph {et~al.}(2022)\citenamefont {Baglio},
  \citenamefont {Duhr}, \citenamefont {Mistlberger},\ and\ \citenamefont
  {Szafron}}]{Baglio:2022wzu}%
  \BibitemOpen
  \bibfield  {author} {\bibinfo {author} {\bibfnamefont {J.}~\bibnamefont
  {Baglio}}, \bibinfo {author} {\bibfnamefont {C.}~\bibnamefont {Duhr}},
  \bibinfo {author} {\bibfnamefont {B.}~\bibnamefont {Mistlberger}},\ and\
  \bibinfo {author} {\bibfnamefont {R.}~\bibnamefont {Szafron}},\ }\href
  {https://doi.org/10.1007/JHEP12(2022)066} {\bibfield  {journal} {\bibinfo
  {journal} {JHEP}\ }\textbf {\bibinfo {volume} {12}},\ \bibinfo {pages}
  {066}},\ \Eprint {https://arxiv.org/abs/2209.06138} {arXiv:2209.06138
  [hep-ph]} \BibitemShut {NoStop}%
\bibitem [{\citenamefont {Ciccolini}\ \emph {et~al.}(2003)\citenamefont
  {Ciccolini}, \citenamefont {Dittmaier},\ and\ \citenamefont
  {Kramer}}]{Ciccolini:2003jy}%
  \BibitemOpen
  \bibfield  {author} {\bibinfo {author} {\bibfnamefont {M.~L.}\ \bibnamefont
  {Ciccolini}}, \bibinfo {author} {\bibfnamefont {S.}~\bibnamefont
  {Dittmaier}},\ and\ \bibinfo {author} {\bibfnamefont {M.}~\bibnamefont
  {Kramer}},\ }\href {https://doi.org/10.1103/PhysRevD.68.073003} {\bibfield
  {journal} {\bibinfo  {journal} {Phys. Rev. D}\ }\textbf {\bibinfo {volume}
  {68}},\ \bibinfo {pages} {073003} (\bibinfo {year} {2003})},\ \Eprint
  {https://arxiv.org/abs/hep-ph/0306234} {arXiv:hep-ph/0306234} \BibitemShut
  {NoStop}%
\bibitem [{\citenamefont {Denner}\ \emph {et~al.}(2012)\citenamefont {Denner},
  \citenamefont {Dittmaier}, \citenamefont {Kallweit},\ and\ \citenamefont
  {Muck}}]{Denner:2011id}%
  \BibitemOpen
  \bibfield  {author} {\bibinfo {author} {\bibfnamefont {A.}~\bibnamefont
  {Denner}}, \bibinfo {author} {\bibfnamefont {S.}~\bibnamefont {Dittmaier}},
  \bibinfo {author} {\bibfnamefont {S.}~\bibnamefont {Kallweit}},\ and\
  \bibinfo {author} {\bibfnamefont {A.}~\bibnamefont {Muck}},\ }\href
  {https://doi.org/10.1007/JHEP03(2012)075} {\bibfield  {journal} {\bibinfo
  {journal} {JHEP}\ }\textbf {\bibinfo {volume} {03}},\ \bibinfo {pages}
  {075}},\ \Eprint {https://arxiv.org/abs/1112.5142} {arXiv:1112.5142 [hep-ph]}
  \BibitemShut {NoStop}%
\bibitem [{\citenamefont {Denner}\ \emph {et~al.}(2015)\citenamefont {Denner},
  \citenamefont {Dittmaier}, \citenamefont {Kallweit},\ and\ \citenamefont
  {M{\"u}ck}}]{Denner:2014cla}%
  \BibitemOpen
  \bibfield  {author} {\bibinfo {author} {\bibfnamefont {A.}~\bibnamefont
  {Denner}}, \bibinfo {author} {\bibfnamefont {S.}~\bibnamefont {Dittmaier}},
  \bibinfo {author} {\bibfnamefont {S.}~\bibnamefont {Kallweit}},\ and\
  \bibinfo {author} {\bibfnamefont {A.}~\bibnamefont {M{\"u}ck}},\ }\href
  {https://doi.org/10.1016/j.cpc.2015.04.021} {\bibfield  {journal} {\bibinfo
  {journal} {Comput. Phys. Commun.}\ }\textbf {\bibinfo {volume} {195}},\
  \bibinfo {pages} {161} (\bibinfo {year} {2015})},\ \Eprint
  {https://arxiv.org/abs/1412.5390} {arXiv:1412.5390 [hep-ph]} \BibitemShut
  {NoStop}%
\bibitem [{\citenamefont {Granata}\ \emph {et~al.}(2017)\citenamefont
  {Granata}, \citenamefont {Lindert}, \citenamefont {Oleari},\ and\
  \citenamefont {Pozzorini}}]{Granata:2017iod}%
  \BibitemOpen
  \bibfield  {author} {\bibinfo {author} {\bibfnamefont {F.}~\bibnamefont
  {Granata}}, \bibinfo {author} {\bibfnamefont {J.~M.}\ \bibnamefont
  {Lindert}}, \bibinfo {author} {\bibfnamefont {C.}~\bibnamefont {Oleari}},\
  and\ \bibinfo {author} {\bibfnamefont {S.}~\bibnamefont {Pozzorini}},\ }\href
  {https://doi.org/10.1007/JHEP09(2017)012} {\bibfield  {journal} {\bibinfo
  {journal} {JHEP}\ }\textbf {\bibinfo {volume} {09}},\ \bibinfo {pages}
  {012}},\ \Eprint {https://arxiv.org/abs/1706.03522} {arXiv:1706.03522
  [hep-ph]} \BibitemShut {NoStop}%
\bibitem [{\citenamefont {Ferrera}\ \emph {et~al.}(2011)\citenamefont
  {Ferrera}, \citenamefont {Grazzini},\ and\ \citenamefont
  {Tramontano}}]{Ferrera:2011bk}%
  \BibitemOpen
  \bibfield  {author} {\bibinfo {author} {\bibfnamefont {G.}~\bibnamefont
  {Ferrera}}, \bibinfo {author} {\bibfnamefont {M.}~\bibnamefont {Grazzini}},\
  and\ \bibinfo {author} {\bibfnamefont {F.}~\bibnamefont {Tramontano}},\
  }\href {https://doi.org/10.1103/PhysRevLett.107.152003} {\bibfield  {journal}
  {\bibinfo  {journal} {Phys. Rev. Lett.}\ }\textbf {\bibinfo {volume} {107}},\
  \bibinfo {pages} {152003} (\bibinfo {year} {2011})},\ \Eprint
  {https://arxiv.org/abs/1107.1164} {arXiv:1107.1164 [hep-ph]} \BibitemShut
  {NoStop}%
\bibitem [{\citenamefont {Ferrera}\ \emph {et~al.}(2014)\citenamefont
  {Ferrera}, \citenamefont {Grazzini},\ and\ \citenamefont
  {Tramontano}}]{Ferrera:2013yga}%
  \BibitemOpen
  \bibfield  {author} {\bibinfo {author} {\bibfnamefont {G.}~\bibnamefont
  {Ferrera}}, \bibinfo {author} {\bibfnamefont {M.}~\bibnamefont {Grazzini}},\
  and\ \bibinfo {author} {\bibfnamefont {F.}~\bibnamefont {Tramontano}},\
  }\href {https://doi.org/10.1007/JHEP04(2014)039} {\bibfield  {journal}
  {\bibinfo  {journal} {JHEP}\ }\textbf {\bibinfo {volume} {04}},\ \bibinfo
  {pages} {039}},\ \Eprint {https://arxiv.org/abs/1312.1669} {arXiv:1312.1669
  [hep-ph]} \BibitemShut {NoStop}%
\bibitem [{\citenamefont {Astill}\ \emph {et~al.}(2016)\citenamefont {Astill},
  \citenamefont {Bizon}, \citenamefont {Re},\ and\ \citenamefont
  {Zanderighi}}]{Astill:2016hpa}%
  \BibitemOpen
  \bibfield  {author} {\bibinfo {author} {\bibfnamefont {W.}~\bibnamefont
  {Astill}}, \bibinfo {author} {\bibfnamefont {W.}~\bibnamefont {Bizon}},
  \bibinfo {author} {\bibfnamefont {E.}~\bibnamefont {Re}},\ and\ \bibinfo
  {author} {\bibfnamefont {G.}~\bibnamefont {Zanderighi}},\ }\href
  {https://doi.org/10.1007/JHEP06(2016)154} {\bibfield  {journal} {\bibinfo
  {journal} {JHEP}\ }\textbf {\bibinfo {volume} {06}},\ \bibinfo {pages}
  {154}},\ \Eprint {https://arxiv.org/abs/1603.01620} {arXiv:1603.01620
  [hep-ph]} \BibitemShut {NoStop}%
\bibitem [{\citenamefont {Campbell}\ \emph {et~al.}(2016)\citenamefont
  {Campbell}, \citenamefont {Ellis},\ and\ \citenamefont
  {Williams}}]{Campbell:2016jau}%
  \BibitemOpen
  \bibfield  {author} {\bibinfo {author} {\bibfnamefont {J.~M.}\ \bibnamefont
  {Campbell}}, \bibinfo {author} {\bibfnamefont {R.~K.}\ \bibnamefont
  {Ellis}},\ and\ \bibinfo {author} {\bibfnamefont {C.}~\bibnamefont
  {Williams}},\ }\href {https://doi.org/10.1007/JHEP06(2016)179} {\bibfield
  {journal} {\bibinfo  {journal} {JHEP}\ }\textbf {\bibinfo {volume} {06}},\
  \bibinfo {pages} {179}},\ \Eprint {https://arxiv.org/abs/1601.00658}
  {arXiv:1601.00658 [hep-ph]} \BibitemShut {NoStop}%
\bibitem [{\citenamefont {Caola}\ \emph {et~al.}(2018)\citenamefont {Caola},
  \citenamefont {Luisoni}, \citenamefont {Melnikov},\ and\ \citenamefont
  {R{\"o}ntsch}}]{Caola:2017xuq}%
  \BibitemOpen
  \bibfield  {author} {\bibinfo {author} {\bibfnamefont {F.}~\bibnamefont
  {Caola}}, \bibinfo {author} {\bibfnamefont {G.}~\bibnamefont {Luisoni}},
  \bibinfo {author} {\bibfnamefont {K.}~\bibnamefont {Melnikov}},\ and\
  \bibinfo {author} {\bibfnamefont {R.}~\bibnamefont {R{\"o}ntsch}},\ }\href
  {https://doi.org/10.1103/PhysRevD.97.074022} {\bibfield  {journal} {\bibinfo
  {journal} {Phys. Rev. D}\ }\textbf {\bibinfo {volume} {97}},\ \bibinfo
  {pages} {074022} (\bibinfo {year} {2018})},\ \Eprint
  {https://arxiv.org/abs/1712.06954} {arXiv:1712.06954 [hep-ph]} \BibitemShut
  {NoStop}%
\bibitem [{\citenamefont {Ferrera}\ \emph {et~al.}(2018)\citenamefont
  {Ferrera}, \citenamefont {Somogyi},\ and\ \citenamefont
  {Tramontano}}]{Ferrera:2017zex}%
  \BibitemOpen
  \bibfield  {author} {\bibinfo {author} {\bibfnamefont {G.}~\bibnamefont
  {Ferrera}}, \bibinfo {author} {\bibfnamefont {G.}~\bibnamefont {Somogyi}},\
  and\ \bibinfo {author} {\bibfnamefont {F.}~\bibnamefont {Tramontano}},\
  }\href {https://doi.org/10.1016/j.physletb.2018.03.021} {\bibfield  {journal}
  {\bibinfo  {journal} {Phys. Lett. B}\ }\textbf {\bibinfo {volume} {780}},\
  \bibinfo {pages} {346} (\bibinfo {year} {2018})},\ \Eprint
  {https://arxiv.org/abs/1705.10304} {arXiv:1705.10304 [hep-ph]} \BibitemShut
  {NoStop}%
\bibitem [{\citenamefont {Alioli}\ \emph {et~al.}(2019)\citenamefont {Alioli},
  \citenamefont {Broggio}, \citenamefont {Kallweit}, \citenamefont {Lim},\ and\
  \citenamefont {Rottoli}}]{Alioli:2019qzz}%
  \BibitemOpen
  \bibfield  {author} {\bibinfo {author} {\bibfnamefont {S.}~\bibnamefont
  {Alioli}}, \bibinfo {author} {\bibfnamefont {A.}~\bibnamefont {Broggio}},
  \bibinfo {author} {\bibfnamefont {S.}~\bibnamefont {Kallweit}}, \bibinfo
  {author} {\bibfnamefont {M.~A.}\ \bibnamefont {Lim}},\ and\ \bibinfo {author}
  {\bibfnamefont {L.}~\bibnamefont {Rottoli}},\ }\href
  {https://doi.org/10.1103/PhysRevD.100.096016} {\bibfield  {journal} {\bibinfo
   {journal} {Phys. Rev. D}\ }\textbf {\bibinfo {volume} {100}},\ \bibinfo
  {pages} {096016} (\bibinfo {year} {2019})},\ \Eprint
  {https://arxiv.org/abs/1909.02026} {arXiv:1909.02026 [hep-ph]} \BibitemShut
  {NoStop}%
\bibitem [{\citenamefont {Behring}\ \emph {et~al.}(2020)\citenamefont
  {Behring}, \citenamefont {Bizo{\'n}}, \citenamefont {Caola}, \citenamefont
  {Melnikov},\ and\ \citenamefont {R{\"o}ntsch}}]{Behring:2020uzq}%
  \BibitemOpen
  \bibfield  {author} {\bibinfo {author} {\bibfnamefont {A.}~\bibnamefont
  {Behring}}, \bibinfo {author} {\bibfnamefont {W.}~\bibnamefont {Bizo{\'n}}},
  \bibinfo {author} {\bibfnamefont {F.}~\bibnamefont {Caola}}, \bibinfo
  {author} {\bibfnamefont {K.}~\bibnamefont {Melnikov}},\ and\ \bibinfo
  {author} {\bibfnamefont {R.}~\bibnamefont {R{\"o}ntsch}},\ }\href
  {https://doi.org/10.1103/PhysRevD.101.114012} {\bibfield  {journal} {\bibinfo
   {journal} {Phys. Rev. D}\ }\textbf {\bibinfo {volume} {101}},\ \bibinfo
  {pages} {114012} (\bibinfo {year} {2020})},\ \Eprint
  {https://arxiv.org/abs/2003.08321} {arXiv:2003.08321 [hep-ph]} \BibitemShut
  {NoStop}%
\bibitem [{\citenamefont {Majer}(2020)}]{Majer:2020kdg}%
  \BibitemOpen
  \bibfield  {author} {\bibinfo {author} {\bibfnamefont {I.}~\bibnamefont
  {Majer}},\ }\emph {\bibinfo {title} {{Associated Higgs Boson Production at
  NNLO QCD}}},\ \href {https://doi.org/10.3929/ethz-b-000448848} {Ph.D.
  thesis},\ \bibinfo  {school} {Zurich, ETH, Zurich, ETH} (\bibinfo {year}
  {2020})\BibitemShut {NoStop}%
\bibitem [{\citenamefont {Zanoli}\ \emph {et~al.}(2022)\citenamefont {Zanoli},
  \citenamefont {Chiesa}, \citenamefont {Re}, \citenamefont {Wiesemann},\ and\
  \citenamefont {Zanderighi}}]{Zanoli:2021iyp}%
  \BibitemOpen
  \bibfield  {author} {\bibinfo {author} {\bibfnamefont {S.}~\bibnamefont
  {Zanoli}}, \bibinfo {author} {\bibfnamefont {M.}~\bibnamefont {Chiesa}},
  \bibinfo {author} {\bibfnamefont {E.}~\bibnamefont {Re}}, \bibinfo {author}
  {\bibfnamefont {M.}~\bibnamefont {Wiesemann}},\ and\ \bibinfo {author}
  {\bibfnamefont {G.}~\bibnamefont {Zanderighi}},\ }\href
  {https://doi.org/10.1007/JHEP07(2022)008} {\bibfield  {journal} {\bibinfo
  {journal} {JHEP}\ }\textbf {\bibinfo {volume} {07}},\ \bibinfo {pages}
  {008}},\ \Eprint {https://arxiv.org/abs/2112.04168} {arXiv:2112.04168
  [hep-ph]} \BibitemShut {NoStop}%
\bibitem [{\citenamefont {Gauld}\ \emph {et~al.}(2019)\citenamefont {Gauld},
  \citenamefont {Gehrmann-De~Ridder}, \citenamefont {Glover}, \citenamefont
  {Huss},\ and\ \citenamefont {Majer}}]{Gauld:2019yng}%
  \BibitemOpen
  \bibfield  {author} {\bibinfo {author} {\bibfnamefont {R.}~\bibnamefont
  {Gauld}}, \bibinfo {author} {\bibfnamefont {A.}~\bibnamefont
  {Gehrmann-De~Ridder}}, \bibinfo {author} {\bibfnamefont {E.~W.~N.}\
  \bibnamefont {Glover}}, \bibinfo {author} {\bibfnamefont {A.}~\bibnamefont
  {Huss}},\ and\ \bibinfo {author} {\bibfnamefont {I.}~\bibnamefont {Majer}},\
  }\href {https://doi.org/10.1007/JHEP10(2019)002} {\bibfield  {journal}
  {\bibinfo  {journal} {JHEP}\ }\textbf {\bibinfo {volume} {10}},\ \bibinfo
  {pages} {002}},\ \Eprint {https://arxiv.org/abs/1907.05836} {arXiv:1907.05836
  [hep-ph]} \BibitemShut {NoStop}%
\bibitem [{\citenamefont {Haisch}\ \emph {et~al.}(2022)\citenamefont {Haisch},
  \citenamefont {Scott}, \citenamefont {Wiesemann}, \citenamefont
  {Zanderighi},\ and\ \citenamefont {Zanoli}}]{Haisch:2022nwz}%
  \BibitemOpen
  \bibfield  {author} {\bibinfo {author} {\bibfnamefont {U.}~\bibnamefont
  {Haisch}}, \bibinfo {author} {\bibfnamefont {D.~J.}\ \bibnamefont {Scott}},
  \bibinfo {author} {\bibfnamefont {M.}~\bibnamefont {Wiesemann}}, \bibinfo
  {author} {\bibfnamefont {G.}~\bibnamefont {Zanderighi}},\ and\ \bibinfo
  {author} {\bibfnamefont {S.}~\bibnamefont {Zanoli}},\ }\href
  {https://doi.org/10.1007/JHEP07(2022)054} {\bibfield  {journal} {\bibinfo
  {journal} {JHEP}\ }\textbf {\bibinfo {volume} {07}},\ \bibinfo {pages}
  {054}},\ \Eprint {https://arxiv.org/abs/2204.00663} {arXiv:2204.00663
  [hep-ph]} \BibitemShut {NoStop}%
\bibitem [{\citenamefont {Anastasiou}\ \emph {et~al.}(2015)\citenamefont
  {Anastasiou}, \citenamefont {Duhr}, \citenamefont {Dulat}, \citenamefont
  {Herzog},\ and\ \citenamefont {Mistlberger}}]{Anastasiou:2015vya}%
  \BibitemOpen
  \bibfield  {author} {\bibinfo {author} {\bibfnamefont {C.}~\bibnamefont
  {Anastasiou}}, \bibinfo {author} {\bibfnamefont {C.}~\bibnamefont {Duhr}},
  \bibinfo {author} {\bibfnamefont {F.}~\bibnamefont {Dulat}}, \bibinfo
  {author} {\bibfnamefont {F.}~\bibnamefont {Herzog}},\ and\ \bibinfo {author}
  {\bibfnamefont {B.}~\bibnamefont {Mistlberger}},\ }\href
  {https://doi.org/10.1103/PhysRevLett.114.212001} {\bibfield  {journal}
  {\bibinfo  {journal} {Phys. Rev. Lett.}\ }\textbf {\bibinfo {volume} {114}},\
  \bibinfo {pages} {212001} (\bibinfo {year} {2015})},\ \Eprint
  {https://arxiv.org/abs/1503.06056} {arXiv:1503.06056 [hep-ph]} \BibitemShut
  {NoStop}%
\bibitem [{\citenamefont {Anastasiou}\ \emph {et~al.}(2016)\citenamefont
  {Anastasiou}, \citenamefont {Duhr}, \citenamefont {Dulat}, \citenamefont
  {Furlan}, \citenamefont {Gehrmann}, \citenamefont {Herzog}, \citenamefont
  {Lazopoulos},\ and\ \citenamefont {Mistlberger}}]{Anastasiou:2016cez}%
  \BibitemOpen
  \bibfield  {author} {\bibinfo {author} {\bibfnamefont {C.}~\bibnamefont
  {Anastasiou}}, \bibinfo {author} {\bibfnamefont {C.}~\bibnamefont {Duhr}},
  \bibinfo {author} {\bibfnamefont {F.}~\bibnamefont {Dulat}}, \bibinfo
  {author} {\bibfnamefont {E.}~\bibnamefont {Furlan}}, \bibinfo {author}
  {\bibfnamefont {T.}~\bibnamefont {Gehrmann}}, \bibinfo {author}
  {\bibfnamefont {F.}~\bibnamefont {Herzog}}, \bibinfo {author} {\bibfnamefont
  {A.}~\bibnamefont {Lazopoulos}},\ and\ \bibinfo {author} {\bibfnamefont
  {B.}~\bibnamefont {Mistlberger}},\ }\href
  {https://doi.org/10.1007/JHEP05(2016)058} {\bibfield  {journal} {\bibinfo
  {journal} {JHEP}\ }\textbf {\bibinfo {volume} {05}},\ \bibinfo {pages}
  {058}},\ \Eprint {https://arxiv.org/abs/1602.00695} {arXiv:1602.00695
  [hep-ph]} \BibitemShut {NoStop}%
\bibitem [{\citenamefont {Dreyer}\ and\ \citenamefont
  {Karlberg}(2016)}]{Dreyer:2016oyx}%
  \BibitemOpen
  \bibfield  {author} {\bibinfo {author} {\bibfnamefont {F.~A.}\ \bibnamefont
  {Dreyer}}\ and\ \bibinfo {author} {\bibfnamefont {A.}~\bibnamefont
  {Karlberg}},\ }\href {https://doi.org/10.1103/PhysRevLett.117.072001}
  {\bibfield  {journal} {\bibinfo  {journal} {Phys. Rev. Lett.}\ }\textbf
  {\bibinfo {volume} {117}},\ \bibinfo {pages} {072001} (\bibinfo {year}
  {2016})},\ \Eprint {https://arxiv.org/abs/1606.00840} {arXiv:1606.00840
  [hep-ph]} \BibitemShut {NoStop}%
\bibitem [{\citenamefont {Mistlberger}(2018)}]{Mistlberger:2018etf}%
  \BibitemOpen
  \bibfield  {author} {\bibinfo {author} {\bibfnamefont {B.}~\bibnamefont
  {Mistlberger}},\ }\href {https://doi.org/10.1007/JHEP05(2018)028} {\bibfield
  {journal} {\bibinfo  {journal} {JHEP}\ }\textbf {\bibinfo {volume} {05}},\
  \bibinfo {pages} {028}},\ \Eprint {https://arxiv.org/abs/1802.00833}
  {arXiv:1802.00833 [hep-ph]} \BibitemShut {NoStop}%
\bibitem [{\citenamefont {Dreyer}\ and\ \citenamefont
  {Karlberg}(2018)}]{Dreyer:2018qbw}%
  \BibitemOpen
  \bibfield  {author} {\bibinfo {author} {\bibfnamefont {F.~A.}\ \bibnamefont
  {Dreyer}}\ and\ \bibinfo {author} {\bibfnamefont {A.}~\bibnamefont
  {Karlberg}},\ }\href {https://doi.org/10.1103/PhysRevD.98.114016} {\bibfield
  {journal} {\bibinfo  {journal} {Phys. Rev. D}\ }\textbf {\bibinfo {volume}
  {98}},\ \bibinfo {pages} {114016} (\bibinfo {year} {2018})},\ \Eprint
  {https://arxiv.org/abs/1811.07906} {arXiv:1811.07906 [hep-ph]} \BibitemShut
  {NoStop}%
\bibitem [{\citenamefont {Cieri}\ \emph {et~al.}(2019)\citenamefont {Cieri},
  \citenamefont {Chen}, \citenamefont {Gehrmann}, \citenamefont {Glover},\ and\
  \citenamefont {Huss}}]{Cieri:2018oms}%
  \BibitemOpen
  \bibfield  {author} {\bibinfo {author} {\bibfnamefont {L.}~\bibnamefont
  {Cieri}}, \bibinfo {author} {\bibfnamefont {X.}~\bibnamefont {Chen}},
  \bibinfo {author} {\bibfnamefont {T.}~\bibnamefont {Gehrmann}}, \bibinfo
  {author} {\bibfnamefont {E.~W.~N.}\ \bibnamefont {Glover}},\ and\ \bibinfo
  {author} {\bibfnamefont {A.}~\bibnamefont {Huss}},\ }\href
  {https://doi.org/10.1007/JHEP02(2019)096} {\bibfield  {journal} {\bibinfo
  {journal} {JHEP}\ }\textbf {\bibinfo {volume} {02}},\ \bibinfo {pages}
  {096}},\ \Eprint {https://arxiv.org/abs/1807.11501} {arXiv:1807.11501
  [hep-ph]} \BibitemShut {NoStop}%
\bibitem [{\citenamefont {Chen}\ \emph
  {et~al.}(2020{\natexlab{a}})\citenamefont {Chen}, \citenamefont {Li},
  \citenamefont {Shao},\ and\ \citenamefont {Wang}}]{Chen:2019lzz}%
  \BibitemOpen
  \bibfield  {author} {\bibinfo {author} {\bibfnamefont {L.-B.}\ \bibnamefont
  {Chen}}, \bibinfo {author} {\bibfnamefont {H.~T.}\ \bibnamefont {Li}},
  \bibinfo {author} {\bibfnamefont {H.-S.}\ \bibnamefont {Shao}},\ and\
  \bibinfo {author} {\bibfnamefont {J.}~\bibnamefont {Wang}},\ }\href
  {https://doi.org/10.1016/j.physletb.2020.135292} {\bibfield  {journal}
  {\bibinfo  {journal} {Phys. Lett. B}\ }\textbf {\bibinfo {volume} {803}},\
  \bibinfo {pages} {135292} (\bibinfo {year} {2020}{\natexlab{a}})},\ \Eprint
  {https://arxiv.org/abs/1909.06808} {arXiv:1909.06808 [hep-ph]} \BibitemShut
  {NoStop}%
\bibitem [{\citenamefont {Chen}\ \emph
  {et~al.}(2020{\natexlab{b}})\citenamefont {Chen}, \citenamefont {Li},
  \citenamefont {Shao},\ and\ \citenamefont {Wang}}]{Chen:2019fhs}%
  \BibitemOpen
  \bibfield  {author} {\bibinfo {author} {\bibfnamefont {L.-B.}\ \bibnamefont
  {Chen}}, \bibinfo {author} {\bibfnamefont {H.~T.}\ \bibnamefont {Li}},
  \bibinfo {author} {\bibfnamefont {H.-S.}\ \bibnamefont {Shao}},\ and\
  \bibinfo {author} {\bibfnamefont {J.}~\bibnamefont {Wang}},\ }\href
  {https://doi.org/10.1007/JHEP03(2020)072} {\bibfield  {journal} {\bibinfo
  {journal} {JHEP}\ }\textbf {\bibinfo {volume} {03}},\ \bibinfo {pages}
  {072}},\ \Eprint {https://arxiv.org/abs/1912.13001} {arXiv:1912.13001
  [hep-ph]} \BibitemShut {NoStop}%
\bibitem [{\citenamefont {Duhr}\ \emph
  {et~al.}(2020{\natexlab{a}})\citenamefont {Duhr}, \citenamefont {Dulat},\
  and\ \citenamefont {Mistlberger}}]{Duhr:2019kwi}%
  \BibitemOpen
  \bibfield  {author} {\bibinfo {author} {\bibfnamefont {C.}~\bibnamefont
  {Duhr}}, \bibinfo {author} {\bibfnamefont {F.}~\bibnamefont {Dulat}},\ and\
  \bibinfo {author} {\bibfnamefont {B.}~\bibnamefont {Mistlberger}},\ }\href
  {https://doi.org/10.1103/PhysRevLett.125.051804} {\bibfield  {journal}
  {\bibinfo  {journal} {Phys. Rev. Lett.}\ }\textbf {\bibinfo {volume} {125}},\
  \bibinfo {pages} {051804} (\bibinfo {year} {2020}{\natexlab{a}})},\ \Eprint
  {https://arxiv.org/abs/1904.09990} {arXiv:1904.09990 [hep-ph]} \BibitemShut
  {NoStop}%
\bibitem [{\citenamefont {Duhr}\ \emph
  {et~al.}(2020{\natexlab{b}})\citenamefont {Duhr}, \citenamefont {Dulat},\
  and\ \citenamefont {Mistlberger}}]{Duhr:2020sdp}%
  \BibitemOpen
  \bibfield  {author} {\bibinfo {author} {\bibfnamefont {C.}~\bibnamefont
  {Duhr}}, \bibinfo {author} {\bibfnamefont {F.}~\bibnamefont {Dulat}},\ and\
  \bibinfo {author} {\bibfnamefont {B.}~\bibnamefont {Mistlberger}},\ }\href
  {https://doi.org/10.1007/JHEP11(2020)143} {\bibfield  {journal} {\bibinfo
  {journal} {JHEP}\ }\textbf {\bibinfo {volume} {11}},\ \bibinfo {pages}
  {143}},\ \Eprint {https://arxiv.org/abs/2007.13313} {arXiv:2007.13313
  [hep-ph]} \BibitemShut {NoStop}%
\bibitem [{\citenamefont {Duhr}\ and\ \citenamefont
  {Mistlberger}(2022)}]{Duhr:2021vwj}%
  \BibitemOpen
  \bibfield  {author} {\bibinfo {author} {\bibfnamefont {C.}~\bibnamefont
  {Duhr}}\ and\ \bibinfo {author} {\bibfnamefont {B.}~\bibnamefont
  {Mistlberger}},\ }\href {https://doi.org/10.1007/JHEP03(2022)116} {\bibfield
  {journal} {\bibinfo  {journal} {JHEP}\ }\textbf {\bibinfo {volume} {03}},\
  \bibinfo {pages} {116}},\ \Eprint {https://arxiv.org/abs/2111.10379}
  {arXiv:2111.10379 [hep-ph]} \BibitemShut {NoStop}%
\bibitem [{\citenamefont {Chen}\ \emph {et~al.}(2021)\citenamefont {Chen},
  \citenamefont {Gehrmann}, \citenamefont {Glover}, \citenamefont {Huss},
  \citenamefont {Mistlberger},\ and\ \citenamefont {Pelloni}}]{Chen:2021isd}%
  \BibitemOpen
  \bibfield  {author} {\bibinfo {author} {\bibfnamefont {X.}~\bibnamefont
  {Chen}}, \bibinfo {author} {\bibfnamefont {T.}~\bibnamefont {Gehrmann}},
  \bibinfo {author} {\bibfnamefont {E.~W.~N.}\ \bibnamefont {Glover}}, \bibinfo
  {author} {\bibfnamefont {A.}~\bibnamefont {Huss}}, \bibinfo {author}
  {\bibfnamefont {B.}~\bibnamefont {Mistlberger}},\ and\ \bibinfo {author}
  {\bibfnamefont {A.}~\bibnamefont {Pelloni}},\ }\href
  {https://doi.org/10.1103/PhysRevLett.127.072002} {\bibfield  {journal}
  {\bibinfo  {journal} {Phys. Rev. Lett.}\ }\textbf {\bibinfo {volume} {127}},\
  \bibinfo {pages} {072002} (\bibinfo {year} {2021})},\ \Eprint
  {https://arxiv.org/abs/2102.07607} {arXiv:2102.07607 [hep-ph]} \BibitemShut
  {NoStop}%
\bibitem [{\citenamefont {Billis}\ \emph {et~al.}(2021)\citenamefont {Billis},
  \citenamefont {Dehnadi}, \citenamefont {Ebert}, \citenamefont {Michel},\ and\
  \citenamefont {Tackmann}}]{Billis:2021ecs}%
  \BibitemOpen
  \bibfield  {author} {\bibinfo {author} {\bibfnamefont {G.}~\bibnamefont
  {Billis}}, \bibinfo {author} {\bibfnamefont {B.}~\bibnamefont {Dehnadi}},
  \bibinfo {author} {\bibfnamefont {M.~A.}\ \bibnamefont {Ebert}}, \bibinfo
  {author} {\bibfnamefont {J.~K.~L.}\ \bibnamefont {Michel}},\ and\ \bibinfo
  {author} {\bibfnamefont {F.~J.}\ \bibnamefont {Tackmann}},\ }\href
  {https://doi.org/10.1103/PhysRevLett.127.072001} {\bibfield  {journal}
  {\bibinfo  {journal} {Phys. Rev. Lett.}\ }\textbf {\bibinfo {volume} {127}},\
  \bibinfo {pages} {072001} (\bibinfo {year} {2021})},\ \Eprint
  {https://arxiv.org/abs/2102.08039} {arXiv:2102.08039 [hep-ph]} \BibitemShut
  {NoStop}%
\bibitem [{\citenamefont {Camarda}\ \emph {et~al.}(2021)\citenamefont
  {Camarda}, \citenamefont {Cieri},\ and\ \citenamefont
  {Ferrera}}]{Camarda:2021ict}%
  \BibitemOpen
  \bibfield  {author} {\bibinfo {author} {\bibfnamefont {S.}~\bibnamefont
  {Camarda}}, \bibinfo {author} {\bibfnamefont {L.}~\bibnamefont {Cieri}},\
  and\ \bibinfo {author} {\bibfnamefont {G.}~\bibnamefont {Ferrera}},\ }\href
  {https://doi.org/10.1103/PhysRevD.104.L111503} {\bibfield  {journal}
  {\bibinfo  {journal} {Phys. Rev. D}\ }\textbf {\bibinfo {volume} {104}},\
  \bibinfo {pages} {L111503} (\bibinfo {year} {2021})},\ \Eprint
  {https://arxiv.org/abs/2103.04974} {arXiv:2103.04974 [hep-ph]} \BibitemShut
  {NoStop}%
\bibitem [{\citenamefont {Chen}\ \emph
  {et~al.}(2022{\natexlab{a}})\citenamefont {Chen}, \citenamefont {Gehrmann},
  \citenamefont {Glover}, \citenamefont {Huss}, \citenamefont {Yang},\ and\
  \citenamefont {Zhu}}]{Chen:2021vtu}%
  \BibitemOpen
  \bibfield  {author} {\bibinfo {author} {\bibfnamefont {X.}~\bibnamefont
  {Chen}}, \bibinfo {author} {\bibfnamefont {T.}~\bibnamefont {Gehrmann}},
  \bibinfo {author} {\bibfnamefont {N.}~\bibnamefont {Glover}}, \bibinfo
  {author} {\bibfnamefont {A.}~\bibnamefont {Huss}}, \bibinfo {author}
  {\bibfnamefont {T.-Z.}\ \bibnamefont {Yang}},\ and\ \bibinfo {author}
  {\bibfnamefont {H.~X.}\ \bibnamefont {Zhu}},\ }\href
  {https://doi.org/10.1103/PhysRevLett.128.052001} {\bibfield  {journal}
  {\bibinfo  {journal} {Phys. Rev. Lett.}\ }\textbf {\bibinfo {volume} {128}},\
  \bibinfo {pages} {052001} (\bibinfo {year} {2022}{\natexlab{a}})},\ \Eprint
  {https://arxiv.org/abs/2107.09085} {arXiv:2107.09085 [hep-ph]} \BibitemShut
  {NoStop}%
\bibitem [{\citenamefont {Chen}\ \emph
  {et~al.}(2022{\natexlab{b}})\citenamefont {Chen}, \citenamefont {Gehrmann},
  \citenamefont {Glover}, \citenamefont {Huss}, \citenamefont {Monni},
  \citenamefont {Re}, \citenamefont {Rottoli},\ and\ \citenamefont
  {Torrielli}}]{Chen:2022cgv}%
  \BibitemOpen
  \bibfield  {author} {\bibinfo {author} {\bibfnamefont {X.}~\bibnamefont
  {Chen}}, \bibinfo {author} {\bibfnamefont {T.}~\bibnamefont {Gehrmann}},
  \bibinfo {author} {\bibfnamefont {E.~W.~N.}\ \bibnamefont {Glover}}, \bibinfo
  {author} {\bibfnamefont {A.}~\bibnamefont {Huss}}, \bibinfo {author}
  {\bibfnamefont {P.~F.}\ \bibnamefont {Monni}}, \bibinfo {author}
  {\bibfnamefont {E.}~\bibnamefont {Re}}, \bibinfo {author} {\bibfnamefont
  {L.}~\bibnamefont {Rottoli}},\ and\ \bibinfo {author} {\bibfnamefont
  {P.}~\bibnamefont {Torrielli}},\ }\href
  {https://doi.org/10.1103/PhysRevLett.128.252001} {\bibfield  {journal}
  {\bibinfo  {journal} {Phys. Rev. Lett.}\ }\textbf {\bibinfo {volume} {128}},\
  \bibinfo {pages} {252001} (\bibinfo {year} {2022}{\natexlab{b}})},\ \Eprint
  {https://arxiv.org/abs/2203.01565} {arXiv:2203.01565 [hep-ph]} \BibitemShut
  {NoStop}%
\bibitem [{\citenamefont {Neumann}\ and\ \citenamefont
  {Campbell}(2023)}]{Neumann:2022lft}%
  \BibitemOpen
  \bibfield  {author} {\bibinfo {author} {\bibfnamefont {T.}~\bibnamefont
  {Neumann}}\ and\ \bibinfo {author} {\bibfnamefont {J.}~\bibnamefont
  {Campbell}},\ }\href {https://doi.org/10.1103/PhysRevD.107.L011506}
  {\bibfield  {journal} {\bibinfo  {journal} {Phys. Rev. D}\ }\textbf {\bibinfo
  {volume} {107}},\ \bibinfo {pages} {L011506} (\bibinfo {year} {2023})},\
  \Eprint {https://arxiv.org/abs/2207.07056} {arXiv:2207.07056 [hep-ph]}
  \BibitemShut {NoStop}%
\bibitem [{\citenamefont {Chen}\ \emph {et~al.}(2023)\citenamefont {Chen},
  \citenamefont {Gehrmann}, \citenamefont {Glover}, \citenamefont {Huss},
  \citenamefont {Yang},\ and\ \citenamefont {Zhu}}]{Chen:2022lwc}%
  \BibitemOpen
  \bibfield  {author} {\bibinfo {author} {\bibfnamefont {X.}~\bibnamefont
  {Chen}}, \bibinfo {author} {\bibfnamefont {T.}~\bibnamefont {Gehrmann}},
  \bibinfo {author} {\bibfnamefont {N.}~\bibnamefont {Glover}}, \bibinfo
  {author} {\bibfnamefont {A.}~\bibnamefont {Huss}}, \bibinfo {author}
  {\bibfnamefont {T.-Z.}\ \bibnamefont {Yang}},\ and\ \bibinfo {author}
  {\bibfnamefont {H.~X.}\ \bibnamefont {Zhu}},\ }\href
  {https://doi.org/10.1016/j.physletb.2023.137876} {\bibfield  {journal}
  {\bibinfo  {journal} {Phys. Lett. B}\ }\textbf {\bibinfo {volume} {840}},\
  \bibinfo {pages} {137876} (\bibinfo {year} {2023})},\ \Eprint
  {https://arxiv.org/abs/2205.11426} {arXiv:2205.11426 [hep-ph]} \BibitemShut
  {NoStop}%
\bibitem [{\citenamefont {Campbell}\ and\ \citenamefont
  {Neumann}(2023)}]{Campbell:2023lcy}%
  \BibitemOpen
  \bibfield  {author} {\bibinfo {author} {\bibfnamefont {J.}~\bibnamefont
  {Campbell}}\ and\ \bibinfo {author} {\bibfnamefont {T.}~\bibnamefont
  {Neumann}},\ }\href {https://doi.org/10.1007/JHEP11(2023)127} {\bibfield
  {journal} {\bibinfo  {journal} {JHEP}\ }\textbf {\bibinfo {volume} {11}},\
  \bibinfo {pages} {127}},\ \Eprint {https://arxiv.org/abs/2308.15382}
  {arXiv:2308.15382 [hep-ph]} \BibitemShut {NoStop}%
\bibitem [{\citenamefont {Czakon}\ \emph {et~al.}(2026)\citenamefont {Czakon},
  \citenamefont {Eschment}, \citenamefont {Generet},\ and\ \citenamefont
  {Poncelet}}]{Czakon:2026zhi}%
  \BibitemOpen
  \bibfield  {author} {\bibinfo {author} {\bibfnamefont {M.}~\bibnamefont
  {Czakon}}, \bibinfo {author} {\bibfnamefont {F.}~\bibnamefont {Eschment}},
  \bibinfo {author} {\bibfnamefont {T.}~\bibnamefont {Generet}},\ and\ \bibinfo
  {author} {\bibfnamefont {R.}~\bibnamefont {Poncelet}},\ }\href@noop {} {\
  (\bibinfo {year} {2026})},\ \Eprint {https://arxiv.org/abs/2604.12613}
  {arXiv:2604.12613 [hep-ph]} \BibitemShut {NoStop}%
\bibitem [{\citenamefont {Chen}\ \emph {et~al.}(2026)\citenamefont {Chen},
  \citenamefont {Dai}, \citenamefont {Li}, \citenamefont {Li}, \citenamefont
  {Shao},\ and\ \citenamefont {Wang}}]{Chen:2026zmi}%
  \BibitemOpen
  \bibfield  {author} {\bibinfo {author} {\bibfnamefont {X.}~\bibnamefont
  {Chen}}, \bibinfo {author} {\bibfnamefont {Y.}~\bibnamefont {Dai}}, \bibinfo
  {author} {\bibfnamefont {H.~T.}\ \bibnamefont {Li}}, \bibinfo {author}
  {\bibfnamefont {S.-Y.}\ \bibnamefont {Li}}, \bibinfo {author} {\bibfnamefont
  {H.-S.}\ \bibnamefont {Shao}},\ and\ \bibinfo {author} {\bibfnamefont
  {J.}~\bibnamefont {Wang}},\ }\href@noop {} {\  (\bibinfo {year} {2026})},\
  \Eprint {https://arxiv.org/abs/2601.19990} {arXiv:2601.19990 [hep-ph]}
  \BibitemShut {NoStop}%
\bibitem [{\citenamefont {Catani}\ and\ \citenamefont
  {Grazzini}(2007)}]{Catani:2007vq}%
  \BibitemOpen
  \bibfield  {author} {\bibinfo {author} {\bibfnamefont {S.}~\bibnamefont
  {Catani}}\ and\ \bibinfo {author} {\bibfnamefont {M.}~\bibnamefont
  {Grazzini}},\ }\href {https://doi.org/10.1103/PhysRevLett.98.222002}
  {\bibfield  {journal} {\bibinfo  {journal} {Phys. Rev. Lett.}\ }\textbf
  {\bibinfo {volume} {98}},\ \bibinfo {pages} {222002} (\bibinfo {year}
  {2007})},\ \Eprint {https://arxiv.org/abs/hep-ph/0703012}
  {arXiv:hep-ph/0703012} \BibitemShut {NoStop}%
\bibitem [{\citenamefont {Gehrmann}\ \emph {et~al.}(2010)\citenamefont
  {Gehrmann}, \citenamefont {Glover}, \citenamefont {Huber}, \citenamefont
  {Ikizlerli},\ and\ \citenamefont {Studerus}}]{Gehrmann:2010ue}%
  \BibitemOpen
  \bibfield  {author} {\bibinfo {author} {\bibfnamefont {T.}~\bibnamefont
  {Gehrmann}}, \bibinfo {author} {\bibfnamefont {E.~W.~N.}\ \bibnamefont
  {Glover}}, \bibinfo {author} {\bibfnamefont {T.}~\bibnamefont {Huber}},
  \bibinfo {author} {\bibfnamefont {N.}~\bibnamefont {Ikizlerli}},\ and\
  \bibinfo {author} {\bibfnamefont {C.}~\bibnamefont {Studerus}},\ }\href
  {https://doi.org/10.1007/JHEP06(2010)094} {\bibfield  {journal} {\bibinfo
  {journal} {JHEP}\ }\textbf {\bibinfo {volume} {06}},\ \bibinfo {pages}
  {094}},\ \Eprint {https://arxiv.org/abs/1004.3653} {arXiv:1004.3653 [hep-ph]}
  \BibitemShut {NoStop}%
\bibitem [{\citenamefont {Catani}\ \emph {et~al.}(2012)\citenamefont {Catani},
  \citenamefont {Cieri}, \citenamefont {de~Florian}, \citenamefont {Ferrera},\
  and\ \citenamefont {Grazzini}}]{Catani:2012qa}%
  \BibitemOpen
  \bibfield  {author} {\bibinfo {author} {\bibfnamefont {S.}~\bibnamefont
  {Catani}}, \bibinfo {author} {\bibfnamefont {L.}~\bibnamefont {Cieri}},
  \bibinfo {author} {\bibfnamefont {D.}~\bibnamefont {de~Florian}}, \bibinfo
  {author} {\bibfnamefont {G.}~\bibnamefont {Ferrera}},\ and\ \bibinfo {author}
  {\bibfnamefont {M.}~\bibnamefont {Grazzini}},\ }\href
  {https://doi.org/10.1140/epjc/s10052-012-2195-7} {\bibfield  {journal}
  {\bibinfo  {journal} {Eur. Phys. J. C}\ }\textbf {\bibinfo {volume} {72}},\
  \bibinfo {pages} {2195} (\bibinfo {year} {2012})},\ \Eprint
  {https://arxiv.org/abs/1209.0158} {arXiv:1209.0158 [hep-ph]} \BibitemShut
  {NoStop}%
\bibitem [{\citenamefont {Gehrmann}\ \emph {et~al.}(2014)\citenamefont
  {Gehrmann}, \citenamefont {Luebbert},\ and\ \citenamefont
  {Yang}}]{Gehrmann:2014yya}%
  \BibitemOpen
  \bibfield  {author} {\bibinfo {author} {\bibfnamefont {T.}~\bibnamefont
  {Gehrmann}}, \bibinfo {author} {\bibfnamefont {T.}~\bibnamefont {Luebbert}},\
  and\ \bibinfo {author} {\bibfnamefont {L.~L.}\ \bibnamefont {Yang}},\ }\href
  {https://doi.org/10.1007/JHEP06(2014)155} {\bibfield  {journal} {\bibinfo
  {journal} {JHEP}\ }\textbf {\bibinfo {volume} {06}},\ \bibinfo {pages}
  {155}},\ \Eprint {https://arxiv.org/abs/1403.6451} {arXiv:1403.6451 [hep-ph]}
  \BibitemShut {NoStop}%
\bibitem [{\citenamefont {L{\"u}bbert}\ \emph {et~al.}(2016)\citenamefont
  {L{\"u}bbert}, \citenamefont {Oredsson},\ and\ \citenamefont
  {Stahlhofen}}]{Lubbert:2016rku}%
  \BibitemOpen
  \bibfield  {author} {\bibinfo {author} {\bibfnamefont {T.}~\bibnamefont
  {L{\"u}bbert}}, \bibinfo {author} {\bibfnamefont {J.}~\bibnamefont
  {Oredsson}},\ and\ \bibinfo {author} {\bibfnamefont {M.}~\bibnamefont
  {Stahlhofen}},\ }\href {https://doi.org/10.1007/JHEP03(2016)168} {\bibfield
  {journal} {\bibinfo  {journal} {JHEP}\ }\textbf {\bibinfo {volume} {03}},\
  \bibinfo {pages} {168}},\ \Eprint {https://arxiv.org/abs/1602.01829}
  {arXiv:1602.01829 [hep-ph]} \BibitemShut {NoStop}%
\bibitem [{\citenamefont {Echevarria}\ \emph {et~al.}(2016)\citenamefont
  {Echevarria}, \citenamefont {Scimemi},\ and\ \citenamefont
  {Vladimirov}}]{Echevarria:2016scs}%
  \BibitemOpen
  \bibfield  {author} {\bibinfo {author} {\bibfnamefont {M.~G.}\ \bibnamefont
  {Echevarria}}, \bibinfo {author} {\bibfnamefont {I.}~\bibnamefont
  {Scimemi}},\ and\ \bibinfo {author} {\bibfnamefont {A.}~\bibnamefont
  {Vladimirov}},\ }\href {https://doi.org/10.1007/JHEP09(2016)004} {\bibfield
  {journal} {\bibinfo  {journal} {JHEP}\ }\textbf {\bibinfo {volume} {09}},\
  \bibinfo {pages} {004}},\ \Eprint {https://arxiv.org/abs/1604.07869}
  {arXiv:1604.07869 [hep-ph]} \BibitemShut {NoStop}%
\bibitem [{\citenamefont {Li}\ and\ \citenamefont {Zhu}(2017)}]{Li:2016ctv}%
  \BibitemOpen
  \bibfield  {author} {\bibinfo {author} {\bibfnamefont {Y.}~\bibnamefont
  {Li}}\ and\ \bibinfo {author} {\bibfnamefont {H.~X.}\ \bibnamefont {Zhu}},\
  }\href {https://doi.org/10.1103/PhysRevLett.118.022004} {\bibfield  {journal}
  {\bibinfo  {journal} {Phys. Rev. Lett.}\ }\textbf {\bibinfo {volume} {118}},\
  \bibinfo {pages} {022004} (\bibinfo {year} {2017})},\ \Eprint
  {https://arxiv.org/abs/1604.01404} {arXiv:1604.01404 [hep-ph]} \BibitemShut
  {NoStop}%
\bibitem [{\citenamefont {Vladimirov}(2017)}]{Vladimirov:2016dll}%
  \BibitemOpen
  \bibfield  {author} {\bibinfo {author} {\bibfnamefont {A.~A.}\ \bibnamefont
  {Vladimirov}},\ }\href {https://doi.org/10.1103/PhysRevLett.118.062001}
  {\bibfield  {journal} {\bibinfo  {journal} {Phys. Rev. Lett.}\ }\textbf
  {\bibinfo {volume} {118}},\ \bibinfo {pages} {062001} (\bibinfo {year}
  {2017})},\ \Eprint {https://arxiv.org/abs/1610.05791} {arXiv:1610.05791
  [hep-ph]} \BibitemShut {NoStop}%
\bibitem [{\citenamefont {Luo}\ \emph {et~al.}(2020)\citenamefont {Luo},
  \citenamefont {Yang}, \citenamefont {Zhu},\ and\ \citenamefont
  {Zhu}}]{Luo:2019szz}%
  \BibitemOpen
  \bibfield  {author} {\bibinfo {author} {\bibfnamefont {M.-x.}\ \bibnamefont
  {Luo}}, \bibinfo {author} {\bibfnamefont {T.-Z.}\ \bibnamefont {Yang}},
  \bibinfo {author} {\bibfnamefont {H.~X.}\ \bibnamefont {Zhu}},\ and\ \bibinfo
  {author} {\bibfnamefont {Y.~J.}\ \bibnamefont {Zhu}},\ }\href
  {https://doi.org/10.1103/PhysRevLett.124.092001} {\bibfield  {journal}
  {\bibinfo  {journal} {Phys. Rev. Lett.}\ }\textbf {\bibinfo {volume} {124}},\
  \bibinfo {pages} {092001} (\bibinfo {year} {2020})},\ \Eprint
  {https://arxiv.org/abs/1912.05778} {arXiv:1912.05778 [hep-ph]} \BibitemShut
  {NoStop}%
\bibitem [{\citenamefont {Ebert}\ \emph {et~al.}(2020)\citenamefont {Ebert},
  \citenamefont {Mistlberger},\ and\ \citenamefont {Vita}}]{Ebert:2020yqt}%
  \BibitemOpen
  \bibfield  {author} {\bibinfo {author} {\bibfnamefont {M.~A.}\ \bibnamefont
  {Ebert}}, \bibinfo {author} {\bibfnamefont {B.}~\bibnamefont {Mistlberger}},\
  and\ \bibinfo {author} {\bibfnamefont {G.}~\bibnamefont {Vita}},\ }\href
  {https://doi.org/10.1007/JHEP09(2020)146} {\bibfield  {journal} {\bibinfo
  {journal} {JHEP}\ }\textbf {\bibinfo {volume} {09}},\ \bibinfo {pages}
  {146}},\ \Eprint {https://arxiv.org/abs/2006.05329} {arXiv:2006.05329
  [hep-ph]} \BibitemShut {NoStop}%
\bibitem [{\citenamefont {Luo}\ \emph {et~al.}(2021)\citenamefont {Luo},
  \citenamefont {Yang}, \citenamefont {Zhu},\ and\ \citenamefont
  {Zhu}}]{Luo:2020epw}%
  \BibitemOpen
  \bibfield  {author} {\bibinfo {author} {\bibfnamefont {M.-x.}\ \bibnamefont
  {Luo}}, \bibinfo {author} {\bibfnamefont {T.-Z.}\ \bibnamefont {Yang}},
  \bibinfo {author} {\bibfnamefont {H.~X.}\ \bibnamefont {Zhu}},\ and\ \bibinfo
  {author} {\bibfnamefont {Y.~J.}\ \bibnamefont {Zhu}},\ }\href
  {https://doi.org/10.1007/JHEP06(2021)115} {\bibfield  {journal} {\bibinfo
  {journal} {JHEP}\ }\textbf {\bibinfo {volume} {06}},\ \bibinfo {pages}
  {115}},\ \Eprint {https://arxiv.org/abs/2012.03256} {arXiv:2012.03256
  [hep-ph]} \BibitemShut {NoStop}%
\bibitem [{\citenamefont {Monni}\ \emph {et~al.}(2016)\citenamefont {Monni},
  \citenamefont {Re},\ and\ \citenamefont {Torrielli}}]{Monni:2016ktx}%
  \BibitemOpen
  \bibfield  {author} {\bibinfo {author} {\bibfnamefont {P.~F.}\ \bibnamefont
  {Monni}}, \bibinfo {author} {\bibfnamefont {E.}~\bibnamefont {Re}},\ and\
  \bibinfo {author} {\bibfnamefont {P.}~\bibnamefont {Torrielli}},\ }\href
  {https://doi.org/10.1103/PhysRevLett.116.242001} {\bibfield  {journal}
  {\bibinfo  {journal} {Phys. Rev. Lett.}\ }\textbf {\bibinfo {volume} {116}},\
  \bibinfo {pages} {242001} (\bibinfo {year} {2016})},\ \Eprint
  {https://arxiv.org/abs/1604.02191} {arXiv:1604.02191 [hep-ph]} \BibitemShut
  {NoStop}%
\bibitem [{\citenamefont {Bizon}\ \emph {et~al.}(2018)\citenamefont {Bizon},
  \citenamefont {Monni}, \citenamefont {Re}, \citenamefont {Rottoli},\ and\
  \citenamefont {Torrielli}}]{Bizon:2017rah}%
  \BibitemOpen
  \bibfield  {author} {\bibinfo {author} {\bibfnamefont {W.}~\bibnamefont
  {Bizon}}, \bibinfo {author} {\bibfnamefont {P.~F.}\ \bibnamefont {Monni}},
  \bibinfo {author} {\bibfnamefont {E.}~\bibnamefont {Re}}, \bibinfo {author}
  {\bibfnamefont {L.}~\bibnamefont {Rottoli}},\ and\ \bibinfo {author}
  {\bibfnamefont {P.}~\bibnamefont {Torrielli}},\ }\href
  {https://doi.org/10.1007/JHEP02(2018)108} {\bibfield  {journal} {\bibinfo
  {journal} {JHEP}\ }\textbf {\bibinfo {volume} {02}},\ \bibinfo {pages}
  {108}},\ \Eprint {https://arxiv.org/abs/1705.09127} {arXiv:1705.09127
  [hep-ph]} \BibitemShut {NoStop}%
\bibitem [{\citenamefont {Re}\ \emph {et~al.}(2021)\citenamefont {Re},
  \citenamefont {Rottoli},\ and\ \citenamefont {Torrielli}}]{Re:2021con}%
  \BibitemOpen
  \bibfield  {author} {\bibinfo {author} {\bibfnamefont {E.}~\bibnamefont
  {Re}}, \bibinfo {author} {\bibfnamefont {L.}~\bibnamefont {Rottoli}},\ and\
  \bibinfo {author} {\bibfnamefont {P.}~\bibnamefont {Torrielli}},\ }\href
  {https://doi.org/10.1007/JHEP09(2021)108} {\bibfield  {journal} {\bibinfo
  {journal} {JHEP}\ }\textbf {\bibinfo {volume} {2109}},\ \bibinfo {pages}
  {108}},\ \Eprint {https://arxiv.org/abs/2104.07509} {arXiv:2104.07509
  [hep-ph]} \BibitemShut {NoStop}%
\bibitem [{\citenamefont {Huss}\ \emph {et~al.}(2026)\citenamefont {Huss} \emph
  {et~al.}}]{NNLOJET:2025rno}%
  \BibitemOpen
  \bibfield  {author} {\bibinfo {author} {\bibfnamefont {A.}~\bibnamefont
  {Huss}} \emph {et~al.} (\bibinfo {collaboration} {NNLOJET}),\ }\href
  {https://doi.org/10.21468/SciPostPhysCodeb.69} {\bibfield  {journal}
  {\bibinfo  {journal} {SciPost Phys. Codeb.}\ }\textbf {\bibinfo {volume}
  {69}},\ \bibinfo {pages} {1} (\bibinfo {year} {2026})},\ \Eprint
  {https://arxiv.org/abs/2503.22804} {arXiv:2503.22804 [hep-ph]} \BibitemShut
  {NoStop}%
\bibitem [{\citenamefont {Gehrmann-De~Ridder}\ \emph
  {et~al.}(2005)\citenamefont {Gehrmann-De~Ridder}, \citenamefont {Gehrmann},\
  and\ \citenamefont {Glover}}]{Gehrmann-DeRidder:2005btv}%
  \BibitemOpen
  \bibfield  {author} {\bibinfo {author} {\bibfnamefont {A.}~\bibnamefont
  {Gehrmann-De~Ridder}}, \bibinfo {author} {\bibfnamefont {T.}~\bibnamefont
  {Gehrmann}},\ and\ \bibinfo {author} {\bibfnamefont {E.~W.~N.}\ \bibnamefont
  {Glover}},\ }\href {https://doi.org/10.1088/1126-6708/2005/09/056} {\bibfield
   {journal} {\bibinfo  {journal} {JHEP}\ }\textbf {\bibinfo {volume} {09}},\
  \bibinfo {pages} {056}},\ \Eprint {https://arxiv.org/abs/hep-ph/0505111}
  {arXiv:hep-ph/0505111} \BibitemShut {NoStop}%
\bibitem [{\citenamefont {Currie}\ \emph {et~al.}(2013)\citenamefont {Currie},
  \citenamefont {Glover},\ and\ \citenamefont {Wells}}]{Currie:2013vh}%
  \BibitemOpen
  \bibfield  {author} {\bibinfo {author} {\bibfnamefont {J.}~\bibnamefont
  {Currie}}, \bibinfo {author} {\bibfnamefont {E.~W.~N.}\ \bibnamefont
  {Glover}},\ and\ \bibinfo {author} {\bibfnamefont {S.}~\bibnamefont
  {Wells}},\ }\href {https://doi.org/10.1007/JHEP04(2013)066} {\bibfield
  {journal} {\bibinfo  {journal} {JHEP}\ }\textbf {\bibinfo {volume} {04}},\
  \bibinfo {pages} {066}},\ \Eprint {https://arxiv.org/abs/1301.4693}
  {arXiv:1301.4693 [hep-ph]} \BibitemShut {NoStop}%
\bibitem [{\citenamefont {Gauld}\ \emph {et~al.}(2021)\citenamefont {Gauld},
  \citenamefont {Gehrmann-De~Ridder}, \citenamefont {Glover}, \citenamefont
  {Huss},\ and\ \citenamefont {Majer}}]{Gauld:2020ced}%
  \BibitemOpen
  \bibfield  {author} {\bibinfo {author} {\bibfnamefont {R.}~\bibnamefont
  {Gauld}}, \bibinfo {author} {\bibfnamefont {A.}~\bibnamefont
  {Gehrmann-De~Ridder}}, \bibinfo {author} {\bibfnamefont {E.~W.~N.}\
  \bibnamefont {Glover}}, \bibinfo {author} {\bibfnamefont {A.}~\bibnamefont
  {Huss}},\ and\ \bibinfo {author} {\bibfnamefont {I.}~\bibnamefont {Majer}},\
  }\href {https://doi.org/10.1016/j.physletb.2021.136335} {\bibfield  {journal}
  {\bibinfo  {journal} {Phys. Lett. B}\ }\textbf {\bibinfo {volume} {817}},\
  \bibinfo {pages} {136335} (\bibinfo {year} {2021})},\ \Eprint
  {https://arxiv.org/abs/2009.14209} {arXiv:2009.14209 [hep-ph]} \BibitemShut
  {NoStop}%
\bibitem [{\citenamefont {Gauld}\ \emph {et~al.}(2022)\citenamefont {Gauld},
  \citenamefont {Gehrmann-De~Ridder}, \citenamefont {Glover}, \citenamefont
  {Huss},\ and\ \citenamefont {Majer}}]{Gauld:2021ule}%
  \BibitemOpen
  \bibfield  {author} {\bibinfo {author} {\bibfnamefont {R.}~\bibnamefont
  {Gauld}}, \bibinfo {author} {\bibfnamefont {A.}~\bibnamefont
  {Gehrmann-De~Ridder}}, \bibinfo {author} {\bibfnamefont {E.~W.~N.}\
  \bibnamefont {Glover}}, \bibinfo {author} {\bibfnamefont {A.}~\bibnamefont
  {Huss}},\ and\ \bibinfo {author} {\bibfnamefont {I.}~\bibnamefont {Majer}},\
  }\href {https://doi.org/10.1007/JHEP03(2022)008} {\bibfield  {journal}
  {\bibinfo  {journal} {JHEP}\ }\textbf {\bibinfo {volume} {03}},\ \bibinfo
  {pages} {008}},\ \Eprint {https://arxiv.org/abs/2110.12992} {arXiv:2110.12992
  [hep-ph]} \BibitemShut {NoStop}%
\bibitem [{\citenamefont {Buccioni}\ \emph {et~al.}(2019)\citenamefont
  {Buccioni}, \citenamefont {Lang}, \citenamefont {Lindert}, \citenamefont
  {Maierh{\"o}fer}, \citenamefont {Pozzorini}, \citenamefont {Zhang},\ and\
  \citenamefont {Zoller}}]{Buccioni:2019sur}%
  \BibitemOpen
  \bibfield  {author} {\bibinfo {author} {\bibfnamefont {F.}~\bibnamefont
  {Buccioni}}, \bibinfo {author} {\bibfnamefont {J.-N.}\ \bibnamefont {Lang}},
  \bibinfo {author} {\bibfnamefont {J.~M.}\ \bibnamefont {Lindert}}, \bibinfo
  {author} {\bibfnamefont {P.}~\bibnamefont {Maierh{\"o}fer}}, \bibinfo
  {author} {\bibfnamefont {S.}~\bibnamefont {Pozzorini}}, \bibinfo {author}
  {\bibfnamefont {H.}~\bibnamefont {Zhang}},\ and\ \bibinfo {author}
  {\bibfnamefont {M.~F.}\ \bibnamefont {Zoller}},\ }\href
  {https://doi.org/10.1140/epjc/s10052-019-7306-2} {\bibfield  {journal}
  {\bibinfo  {journal} {Eur. Phys. J. C}\ }\textbf {\bibinfo {volume} {79}},\
  \bibinfo {pages} {866} (\bibinfo {year} {2019})},\ \Eprint
  {https://arxiv.org/abs/1907.13071} {arXiv:1907.13071 [hep-ph]} \BibitemShut
  {NoStop}%
\bibitem [{\citenamefont {Brein}\ \emph {et~al.}(2012)\citenamefont {Brein},
  \citenamefont {Harlander}, \citenamefont {Wiesemann},\ and\ \citenamefont
  {Zirke}}]{Brein:2011vx}%
  \BibitemOpen
  \bibfield  {author} {\bibinfo {author} {\bibfnamefont {O.}~\bibnamefont
  {Brein}}, \bibinfo {author} {\bibfnamefont {R.}~\bibnamefont {Harlander}},
  \bibinfo {author} {\bibfnamefont {M.}~\bibnamefont {Wiesemann}},\ and\
  \bibinfo {author} {\bibfnamefont {T.}~\bibnamefont {Zirke}},\ }\href
  {https://doi.org/10.1140/epjc/s10052-012-1868-6} {\bibfield  {journal}
  {\bibinfo  {journal} {Eur. Phys. J. C}\ }\textbf {\bibinfo {volume} {72}},\
  \bibinfo {pages} {1868} (\bibinfo {year} {2012})},\ \Eprint
  {https://arxiv.org/abs/1111.0761} {arXiv:1111.0761 [hep-ph]} \BibitemShut
  {NoStop}%
\bibitem [{\citenamefont {Ball}\ \emph {et~al.}(2024)\citenamefont {Ball} \emph
  {et~al.}}]{NNPDF:2024nan}%
  \BibitemOpen
  \bibfield  {author} {\bibinfo {author} {\bibfnamefont {R.~D.}\ \bibnamefont
  {Ball}} \emph {et~al.} (\bibinfo {collaboration} {NNPDF}),\ }\href
  {https://doi.org/10.1140/epjc/s10052-024-12891-7} {\bibfield  {journal}
  {\bibinfo  {journal} {Eur. Phys. J. C}\ }\textbf {\bibinfo {volume} {84}},\
  \bibinfo {pages} {659} (\bibinfo {year} {2024})},\ \Eprint
  {https://arxiv.org/abs/2402.18635} {arXiv:2402.18635 [hep-ph]} \BibitemShut
  {NoStop}%
\bibitem [{\citenamefont {Buckley}\ \emph {et~al.}(2015)\citenamefont
  {Buckley}, \citenamefont {Ferrando}, \citenamefont {Lloyd}, \citenamefont
  {Nordstr{\"o}m}, \citenamefont {Page}, \citenamefont {R{\"u}fenacht},
  \citenamefont {Sch{\"o}nherr},\ and\ \citenamefont {Watt}}]{Buckley:2014ana}%
  \BibitemOpen
  \bibfield  {author} {\bibinfo {author} {\bibfnamefont {A.}~\bibnamefont
  {Buckley}}, \bibinfo {author} {\bibfnamefont {J.}~\bibnamefont {Ferrando}},
  \bibinfo {author} {\bibfnamefont {S.}~\bibnamefont {Lloyd}}, \bibinfo
  {author} {\bibfnamefont {K.}~\bibnamefont {Nordstr{\"o}m}}, \bibinfo {author}
  {\bibfnamefont {B.}~\bibnamefont {Page}}, \bibinfo {author} {\bibfnamefont
  {M.}~\bibnamefont {R{\"u}fenacht}}, \bibinfo {author} {\bibfnamefont
  {M.}~\bibnamefont {Sch{\"o}nherr}},\ and\ \bibinfo {author} {\bibfnamefont
  {G.}~\bibnamefont {Watt}},\ }\href
  {https://doi.org/10.1140/epjc/s10052-015-3318-8} {\bibfield  {journal}
  {\bibinfo  {journal} {Eur. Phys. J. C}\ }\textbf {\bibinfo {volume} {75}},\
  \bibinfo {pages} {132} (\bibinfo {year} {2015})},\ \Eprint
  {https://arxiv.org/abs/1412.7420} {arXiv:1412.7420 [hep-ph]} \BibitemShut
  {NoStop}%
\bibitem [{\citenamefont {LHCHXSWG}(2025)}]{LHCHXSWG-RUNIII}%
  \BibitemOpen
  \bibfield  {author} {\bibinfo {author} {\bibnamefont {LHCHXSWG}},\ }\href
  {https://twiki.cern.ch/twiki/bin/view/LHCPhysics/LHCHWG136TeVxsec} {\bibinfo
  {title} {Higgs production cross section for run iii}} (\bibinfo {year}
  {2025})\BibitemShut {NoStop}%
\bibitem [{\citenamefont {Andersen}\ \emph {et~al.}(2016)\citenamefont
  {Andersen} \emph {et~al.}}]{Andersen:2016qtm}%
  \BibitemOpen
  \bibfield  {author} {\bibinfo {author} {\bibfnamefont {J.~R.}\ \bibnamefont
  {Andersen}} \emph {et~al.},\ }in\ \href@noop {} {\emph {\bibinfo {booktitle}
  {{9th Les Houches Workshop on Physics at TeV Colliders}}}}\ (\bibinfo {year}
  {2016})\ \Eprint {https://arxiv.org/abs/1605.04692} {arXiv:1605.04692
  [hep-ph]} \BibitemShut {NoStop}%
\bibitem [{\citenamefont {Berger}\ \emph {et~al.}(2019)\citenamefont {Berger}
  \emph {et~al.}}]{Berger:2019wnu}%
  \BibitemOpen
  \bibfield  {author} {\bibinfo {author} {\bibfnamefont {N.}~\bibnamefont
  {Berger}} \emph {et~al.},\ }\bibfield  {journal} {\bibinfo  {journal}
  {SciPost Phys. Comm. Rep.}\ }\href
  {https://doi.org/10.21468/SciPostPhysCommRep.15}
  {10.21468/SciPostPhysCommRep.15} (\bibinfo {year} {2019}),\ \Eprint
  {https://arxiv.org/abs/1906.02754} {arXiv:1906.02754 [hep-ph]} \BibitemShut
  {NoStop}%
\bibitem [{\citenamefont {Aaboud}\ \emph {et~al.}(2019)\citenamefont {Aaboud}
  \emph {et~al.}}]{ATLAS:2019yhn}%
  \BibitemOpen
  \bibfield  {author} {\bibinfo {author} {\bibfnamefont {M.}~\bibnamefont
  {Aaboud}} \emph {et~al.} (\bibinfo {collaboration} {ATLAS}),\ }\href
  {https://doi.org/10.1007/JHEP05(2019)141} {\bibfield  {journal} {\bibinfo
  {journal} {JHEP}\ }\textbf {\bibinfo {volume} {05}},\ \bibinfo {pages}
  {141}},\ \Eprint {https://arxiv.org/abs/1903.04618} {arXiv:1903.04618
  [hep-ex]} \BibitemShut {NoStop}%
\bibitem [{\citenamefont {Gehrmann-De~Ridder}\ \emph
  {et~al.}(2026)\citenamefont {Gehrmann-De~Ridder}, \citenamefont {Huss},
  \citenamefont {Marcoli}, \citenamefont {Monni}, \citenamefont {Re},
  \citenamefont {Rottoli}, \citenamefont {Silvetti},\ and\ \citenamefont
  {Torrielli}}]{supplemental}%
  \BibitemOpen
  \bibfield  {author} {\bibinfo {author} {\bibfnamefont {A.}~\bibnamefont
  {Gehrmann-De~Ridder}}, \bibinfo {author} {\bibfnamefont {A.}~\bibnamefont
  {Huss}}, \bibinfo {author} {\bibfnamefont {M.}~\bibnamefont {Marcoli}},
  \bibinfo {author} {\bibfnamefont {P.~F.}\ \bibnamefont {Monni}}, \bibinfo
  {author} {\bibfnamefont {E.}~\bibnamefont {Re}}, \bibinfo {author}
  {\bibfnamefont {L.}~\bibnamefont {Rottoli}}, \bibinfo {author} {\bibfnamefont
  {F.}~\bibnamefont {Silvetti}},\ and\ \bibinfo {author} {\bibfnamefont
  {P.}~\bibnamefont {Torrielli}},\ }\href@noop {} {\  (\bibinfo {year}
  {2026})},\ \Eprint {https://arxiv.org/abs/Supplemental Material to this
  Letter} {Supplemental Material to this Letter} \BibitemShut {NoStop}%
\bibitem [{\citenamefont {Frixione}\ and\ \citenamefont
  {Ridolfi}(1996)}]{Frixione:1996ym}%
  \BibitemOpen
  \bibfield  {author} {\bibinfo {author} {\bibfnamefont {S.}~\bibnamefont
  {Frixione}}\ and\ \bibinfo {author} {\bibfnamefont {G.}~\bibnamefont
  {Ridolfi}},\ }\href {https://doi.org/10.1016/0370-2693(96)00732-0} {\bibfield
   {journal} {\bibinfo  {journal} {Phys. Lett. B}\ }\textbf {\bibinfo {volume}
  {383}},\ \bibinfo {pages} {227} (\bibinfo {year} {1996})},\ \Eprint
  {https://arxiv.org/abs/hep-ph/9605209} {arXiv:hep-ph/9605209} \BibitemShut
  {NoStop}%
\bibitem [{\citenamefont {Grazzini}\ \emph {et~al.}(2018)\citenamefont
  {Grazzini}, \citenamefont {Kallweit},\ and\ \citenamefont
  {Wiesemann}}]{Grazzini:2017mhc}%
  \BibitemOpen
  \bibfield  {author} {\bibinfo {author} {\bibfnamefont {M.}~\bibnamefont
  {Grazzini}}, \bibinfo {author} {\bibfnamefont {S.}~\bibnamefont {Kallweit}},\
  and\ \bibinfo {author} {\bibfnamefont {M.}~\bibnamefont {Wiesemann}},\ }\href
  {https://doi.org/10.1140/epjc/s10052-018-5771-7} {\bibfield  {journal}
  {\bibinfo  {journal} {Eur. Phys. J. C}\ }\textbf {\bibinfo {volume} {78}},\
  \bibinfo {pages} {537} (\bibinfo {year} {2018})},\ \Eprint
  {https://arxiv.org/abs/1711.06631} {arXiv:1711.06631 [hep-ph]} \BibitemShut
  {NoStop}%
\bibitem [{\citenamefont {Ebert}\ and\ \citenamefont
  {Tackmann}(2020)}]{Ebert:2019zkb}%
  \BibitemOpen
  \bibfield  {author} {\bibinfo {author} {\bibfnamefont {M.~A.}\ \bibnamefont
  {Ebert}}\ and\ \bibinfo {author} {\bibfnamefont {F.~J.}\ \bibnamefont
  {Tackmann}},\ }\href {https://doi.org/10.1007/JHEP03(2020)158} {\bibfield
  {journal} {\bibinfo  {journal} {JHEP}\ }\textbf {\bibinfo {volume} {03}},\
  \bibinfo {pages} {158}},\ \Eprint {https://arxiv.org/abs/1911.08486}
  {arXiv:1911.08486 [hep-ph]} \BibitemShut {NoStop}%
\bibitem [{\citenamefont {Salam}\ and\ \citenamefont
  {Slade}(2021)}]{Salam:2021tbm}%
  \BibitemOpen
  \bibfield  {author} {\bibinfo {author} {\bibfnamefont {G.~P.}\ \bibnamefont
  {Salam}}\ and\ \bibinfo {author} {\bibfnamefont {E.}~\bibnamefont {Slade}},\
  }\href {https://doi.org/10.1007/JHEP11(2021)220} {\bibfield  {journal}
  {\bibinfo  {journal} {JHEP}\ }\textbf {\bibinfo {volume} {11}},\ \bibinfo
  {pages} {220}},\ \Eprint {https://arxiv.org/abs/2106.08329} {arXiv:2106.08329
  [hep-ph]} \BibitemShut {NoStop}%
\bibitem [{\citenamefont {Alekhin}\ \emph {et~al.}(2021)\citenamefont
  {Alekhin}, \citenamefont {Kardos}, \citenamefont {Moch},\ and\ \citenamefont
  {Tr{\'o}cs{\'a}nyi}}]{Alekhin:2021xcu}%
  \BibitemOpen
  \bibfield  {author} {\bibinfo {author} {\bibfnamefont {S.}~\bibnamefont
  {Alekhin}}, \bibinfo {author} {\bibfnamefont {A.}~\bibnamefont {Kardos}},
  \bibinfo {author} {\bibfnamefont {S.}~\bibnamefont {Moch}},\ and\ \bibinfo
  {author} {\bibfnamefont {Z.}~\bibnamefont {Tr{\'o}cs{\'a}nyi}},\ }\href
  {https://doi.org/10.1140/epjc/s10052-021-09361-9} {\bibfield  {journal}
  {\bibinfo  {journal} {Eur. Phys. J. C}\ }\textbf {\bibinfo {volume} {81}},\
  \bibinfo {pages} {573} (\bibinfo {year} {2021})},\ \Eprint
  {https://arxiv.org/abs/2104.02400} {arXiv:2104.02400 [hep-ph]} \BibitemShut
  {NoStop}%
\bibitem [{\citenamefont {Buonocore}\ \emph {et~al.}(2022)\citenamefont
  {Buonocore}, \citenamefont {Kallweit}, \citenamefont {Rottoli},\ and\
  \citenamefont {Wiesemann}}]{Buonocore:2021tke}%
  \BibitemOpen
  \bibfield  {author} {\bibinfo {author} {\bibfnamefont {L.}~\bibnamefont
  {Buonocore}}, \bibinfo {author} {\bibfnamefont {S.}~\bibnamefont {Kallweit}},
  \bibinfo {author} {\bibfnamefont {L.}~\bibnamefont {Rottoli}},\ and\ \bibinfo
  {author} {\bibfnamefont {M.}~\bibnamefont {Wiesemann}},\ }\href
  {https://doi.org/10.1016/j.physletb.2022.137118} {\bibfield  {journal}
  {\bibinfo  {journal} {Phys. Lett. B}\ }\textbf {\bibinfo {volume} {829}},\
  \bibinfo {pages} {137118} (\bibinfo {year} {2022})},\ \Eprint
  {https://arxiv.org/abs/2111.13661} {arXiv:2111.13661 [hep-ph]} \BibitemShut
  {NoStop}%
\bibitem [{\citenamefont {Camarda}\ \emph {et~al.}(2022)\citenamefont
  {Camarda}, \citenamefont {Cieri},\ and\ \citenamefont
  {Ferrera}}]{Camarda:2021jsw}%
  \BibitemOpen
  \bibfield  {author} {\bibinfo {author} {\bibfnamefont {S.}~\bibnamefont
  {Camarda}}, \bibinfo {author} {\bibfnamefont {L.}~\bibnamefont {Cieri}},\
  and\ \bibinfo {author} {\bibfnamefont {G.}~\bibnamefont {Ferrera}},\ }\href
  {https://doi.org/10.1140/epjc/s10052-022-10510-x} {\bibfield  {journal}
  {\bibinfo  {journal} {Eur. Phys. J. C}\ }\textbf {\bibinfo {volume} {82}},\
  \bibinfo {pages} {575} (\bibinfo {year} {2022})},\ \Eprint
  {https://arxiv.org/abs/2111.14509} {arXiv:2111.14509 [hep-ph]} \BibitemShut
  {NoStop}%
\bibitem [{\citenamefont {Klasen}\ and\ \citenamefont
  {Kramer}(1996)}]{Klasen:1995xe}%
  \BibitemOpen
  \bibfield  {author} {\bibinfo {author} {\bibfnamefont {M.}~\bibnamefont
  {Klasen}}\ and\ \bibinfo {author} {\bibfnamefont {G.}~\bibnamefont
  {Kramer}},\ }\href {https://doi.org/10.1016/0370-2693(95)01352-0} {\bibfield
  {journal} {\bibinfo  {journal} {Phys. Lett. B}\ }\textbf {\bibinfo {volume}
  {366}},\ \bibinfo {pages} {385} (\bibinfo {year} {1996})},\ \Eprint
  {https://arxiv.org/abs/hep-ph/9508337} {arXiv:hep-ph/9508337} \BibitemShut
  {NoStop}%
\bibitem [{\citenamefont {Harris}\ and\ \citenamefont
  {Owens}(1997)}]{Harris:1997hz}%
  \BibitemOpen
  \bibfield  {author} {\bibinfo {author} {\bibfnamefont {B.~W.}\ \bibnamefont
  {Harris}}\ and\ \bibinfo {author} {\bibfnamefont {J.~F.}\ \bibnamefont
  {Owens}},\ }\href {https://doi.org/10.1103/PhysRevD.56.4007} {\bibfield
  {journal} {\bibinfo  {journal} {Phys. Rev. D}\ }\textbf {\bibinfo {volume}
  {56}},\ \bibinfo {pages} {4007} (\bibinfo {year} {1997})},\ \Eprint
  {https://arxiv.org/abs/hep-ph/9704324} {arXiv:hep-ph/9704324} \BibitemShut
  {NoStop}%
\bibitem [{\citenamefont {Frixione}\ and\ \citenamefont
  {Ridolfi}(1997)}]{Frixione:1997ks}%
  \BibitemOpen
  \bibfield  {author} {\bibinfo {author} {\bibfnamefont {S.}~\bibnamefont
  {Frixione}}\ and\ \bibinfo {author} {\bibfnamefont {G.}~\bibnamefont
  {Ridolfi}},\ }\href {https://doi.org/10.1016/S0550-3213(97)00575-0}
  {\bibfield  {journal} {\bibinfo  {journal} {Nucl. Phys. B}\ }\textbf
  {\bibinfo {volume} {507}},\ \bibinfo {pages} {315} (\bibinfo {year}
  {1997})},\ \Eprint {https://arxiv.org/abs/hep-ph/9707345}
  {arXiv:hep-ph/9707345} \BibitemShut {NoStop}%
\bibitem [{\citenamefont {Karlberg}\ \emph {et~al.}(2024)\citenamefont
  {Karlberg} \emph {et~al.}}]{Karlberg:2024zxx}%
  \BibitemOpen
  \bibfield  {author} {\bibinfo {author} {\bibfnamefont {A.}~\bibnamefont
  {Karlberg}} \emph {et~al.},\ }\href@noop {} {\  (\bibinfo {year} {2024})},\
  \Eprint {https://arxiv.org/abs/2402.09955} {arXiv:2402.09955 [hep-ph]}
  \BibitemShut {NoStop}%
\end{thebibliography}%
